\pdfoutput=1
\documentclass[preprints,article,accept,moreauthors]{Definitions/mdpi}
\usepackage[utf8]{inputenc}
\firstpage{1} 
\pubvolume{1}
\issuenum{1}
\articlenumber{0}
\pubyear{2026}
\copyrightyear{2026}
\datereceived{ } 
\daterevised{ } % Comment out if no revised date
\dateaccepted{ } 
\datepublished{ } 

\usepackage{pdflscape}
\usepackage{rotating}
\usepackage{longtable}
\usepackage{lscape}
\usepackage{longtable}
\usepackage{makecell}
\Title{Solar extreme–ultraviolet waves and other associated solar activities}

\Author{Apoorv Dashora $^{1}$\orcidA{}, Ramesh Chandra $^{1,}$*\orcidB{} and Pooja Devi $^{2,3}$\orcidC{}}

\AuthorNames{ Apoorv Dashora, Ramesh Chandra  and Pooja Devi }

\address{%
$^{1}$ \quad Department of Physics, DSB Campus, Kumaun University, Nainital 263001, India\\
$^{2}$ \quad Rosseland Centre for Solar Physics, University of Oslo, P.O. Box 1029, Blindern, N-0315 Oslo, Norway\\
$^{3}$ \quad Institute of Theoretical Astrophysics, University of Oslo, P.O. Box 1029, Blindern, N-0315 Oslo, Norway\\}

\corres{Correspondence: rchandra.ntl@gmail.com}

\abstract{Extreme ultraviolet (EUV) waves are large-scale, tsunami like disturbances on the solar surface associated with eruptive phenomena such as solar flares, filament eruptions, and coronal mass ejections (CMEs). Because their strong correlation with major solar eruptions, we present here a comprehensive statistical study of EUV waves and their connections to Moreton waves, CMEs, solar flares, type II radio bursts, and solar energetic particle (SEP) events during Solar Cycle 25  (December 2019--December 2025).
This study utilizes EUV data from the Atmospheric Imaging Assembly onboard the Solar Dynamics Observatory and coronagraph data from the Large Angle and Spectrometric Coronagraph onboard Solar and Heliospheric Observatory. Our  analysis reveals that 12.3\% of the observed EUV waves are associated with Moreton waves, 95 \% with CMEs,  92.7\% associated with flares, 44.3 \% with metric type II radio bursts, and 11.8 \% with SEPs. While nearly all EUV waves are linked to CMEs, their association with solar flares varies significantly, ranging from weak B-class events to major X-class flares. Furthermore, the majority of Moreton waves correlate with large- or medium-class flares, with only a small fraction associated with weaker C-class flares. 
Our findings may contribute to a better understanding of space weather events and, potentially, to their forecasting.}

\keyword{Solar EUV Waves; Solar Corona; CMEs ; Solar Flares; Type II radio bursts; SEPs}
\DeclareUnicodeCharacter{03B1}{\ensuremath{\alpha}}
\begin{document}

%\section{Introduction}
\section{Introduction}
\label{sec:intro}

%\noindent{\S}{\bf --- Definition:} 
Solar Extreme-ultraviolet (EUV) waves are globally propagating, wave like perturbations in the solar corona that are observed in the EUV  wavelengths. They are also visible in other wavelengths, including soft X-ray, H$\alpha$ and radio. These spectacular perturbations propagate across solar disk with speeds ranging from a few to more than a thousand km s$^{-1}$.

Prior to detection of EUV waves in corona, their arc-shaped chromospheric counterparts were imaged in H$\alpha$ line centre, featured as bright front in blue wing and as dark front in red wing, and were named Moreton waves \citep{Moreton60, MoretonR60, Athaymoreton1961}. After three and half decades from the discovery of Moreton waves, these enhanced wave like signatures reported during the study of an Earth-directed coronal mass ejection (CME) event of 1997 May 12 with the Extreme-ultraviolet Imaging Telescope \citep[EIT;][]{Delaboudiniere95} aboard the Solar and Heliospheric Observatory \citep[SOHO;][]{Domingo95}. The diffuse and nearly isotropic EUV wave fronts propagated outward from the source location of the erupting flare and were thereafter termed `EIT waves' \citep{Moses1997,Thompson98}. 
Later, these features were called as EUV waves, coronal waves, large-scale propagating fronts, etc. In this paper, we use the name 'EUV waves' throughout to represent wavelike disturbances in EUV wavelength.
\\
\begin{figure*}[!t]  %%%%% Figure 1
\centering
\includegraphics[width=0.99\textwidth]{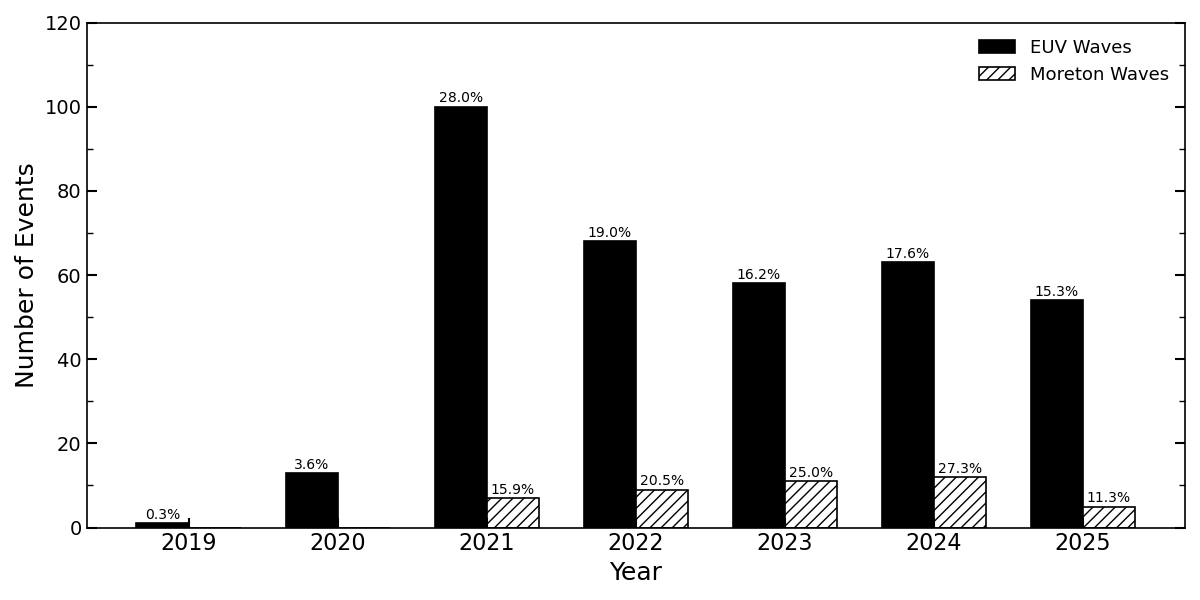}
\caption{Year wise occurrence of EUV waves (black bar) and Moreton waves (hatched bar). }
\label{Year}
\end{figure*}

EUV waves are associated with other solar activities such as  filament eruptions, CMEs \citep{Chen2011,Chandra2016}, flares \citep{Alexander2006}, solar energetic particles (SEPs) \citep{miteva2014}, type II radio burst \citep{Nitta_2013,Chandra2018typeII}.
Studies suggested that EUV waves possess more association with CMEs than Solar flares \citep{Biesecker2002, Chen2006, Nitta2014, Muhr2014,Chandra2024a}.
The study of \cite{Biesecker2002} reported the unambiguous correlation between EUV waves and CMEs. \citet{Chen2006} presented a study of fourteen CME-less solar flares and found that none of the flares were accompanied by EUV waves. Their findings demonstrate that a CME is a necessary condition for the generation of EUV waves. 
There are many studies which show the association of EUV waves with CMEs \citep{ Chen2006, Devi22, Kwon_2014, Kwon_2017}. All these studies use the observations from SDO, STEREO, and LASCO to see the evolution of EUV wave and CME and they found that the fast-component EUV wave is the shock travelling ahead of the CME. Furthermore, \citep{ Kwon_2014,Kwon_2017} show this result to be true using the geometric models. Therefore, the fast-component EUV waves are the shocks driven by CMEs.
\citep{Chen2006, Devi22} and \cite{Chandra2024} presented the case studies to verify that the EUV waves are directly linked with the CMEs, even the fast-component EUV waves are driven by the CMEs.

To explore the association between EUV waves and SEPs, \cite{miteva2014} selected 179 SEP events during Solar Cycle 23 (1997 - 2006), and reported 87\% of SEPs associated with EUV waves. 
Similar to this, the type II radio bursts are another important phenomenon associated with EUV waves. Type II radio bursts are plasma emissions characterized by a slow drift toward lower frequencies in dynamic spectra (plots of radio intensity as a function of time and frequency). Their drift rate varies from 0.1 to 1 MHz s $^{-1} $ in the metric domain, and their lifetime is typically tens of minutes \citep{Nelson_1985}. When the type II radio bursts are observed in the decametric hectometric range, we call them DH type II radio bursts. These radio bursts are produced in the interplanetary medium. The type II radio bursts are signatures of coronal shock waves.
Since the fast-component EUV waves and type II radio bursts are different signatures of the same shock,  therefore a strong association the fast-component between EUV waves and type II radio bursts is highly anticipated. 
To address this relationship, various studies have been performed \citep{Klassen2000, Warmuth10, Nitta_2013, Nitta2014, Long17}. One of these studies by \cite{Klassen2000} found 90 \% of type II radio burst associated with EUV waves. However, some studies suggest their lower association rates with type II radio bursts \citep{Nitta_2013,Muhr14,Chandra2026}. 
 
In all above mentioned phenomena, particularly ICMEs (Interplanetary coronal mass ejections) disturb magnetosphere of earth and fluctuates horizontal intensity of earth magnetic field abruptly. It causes geomagnetic storm (GS) \citep{Gonzalez1994}. The hourly value of this low-latitude magnetic field intensity is called disturbance storm time (Dst) index. In the past, many studies have been performed to analyze relationship between CMEs, ICMEs and GSs characteristics.\cite{Richardson_2011,Syed2019,Samuel_2023,Mitewa2023,MitewaS_2023} 
%\cite{Mitewa2023} analyzed all major GSs of SC 24 and performed correlations between the various parameters of coronal and interplanetary medium.\\

Various models have been proposed to explain the nature of EUV waves. These includes: wave models, non-wave models and hybrid models.
%The physical nature Their are several models The physical nature of EUV waves believed as Fast-mode wave, slow-mode waves and magnetoacoustic wave \citep{Ballai2011}. 
In this context, \cite{Uchida1968} considered counterparts of EUV waves i.e., Moreton waves as fast mode waves and proposed that the dome-shaped wavefronts of Moreton waves extends up to corona and their flanks sweep faster over the chromosphere. 
However, the speed of EUV wavefronts shows discrepancy from fast-mode assumption and decreases below coronal sound speed ($ < 185 $ km s ${^{-1}}$). Therefore, \cite{Wills2009} explained these extended wavefronts as slow-mode via slow-mode soliton model. To describe the propagation of
EUV waves in the low corona, \citep{Ballai2011} described EUV wave as magnetoacoustic surface gravity wave.  
Interestingly, EUV waves are not considered only as real MHD waves. \cite{Delanee1999} found presence of stationary bright front and dimming in three events near magnetic separatrix. 
\cite{Delanne2008} developed current shell model, demonstrating that this brightening can be due to joule heating by rotation of a current shell.  
Similarly, \cite{Attrill2007} presented that the interaction of sequential tiny magnetic loops with laterally expanding CME appears as chain-like propagation of brightening by reconnection front model. Overall, non wave models treat EUV wave as a pseudo wave which is outcome due to rearrangement of magnetic field lines during an erupting CME. To solve this puzzle of wave and non-wave nature for EUV waves, \cite{chen2002,Chen2005b} identified presence of both wave and non-wave components and proposed as `hybrid model'. In their MHD simulations, there are two types of EUV waves; a fast component which is the coronal shock wave due to piston-driven mechanism straddles over flux rope (FR) and the slower wave is explained by magnetic field line stretching model \citep{chen2005a}. According to magnetic field line stretching model, the eruption of flux rope stretches magnetic field lines. This deformation travels successively from inner to outer magnetic field lines and the resultant brightening becomes visible due to compression of plasma.
Hybrid model is currently widely accepted as it combines both wave and non-wave characteristics of EUV waves.\\
 %\noindent{\S}{\bf --- Problem and aim of the paper:} 
\begin{figure*}[!t]  %%%%% Figure 2
\centering
\includegraphics[width=1\textwidth]{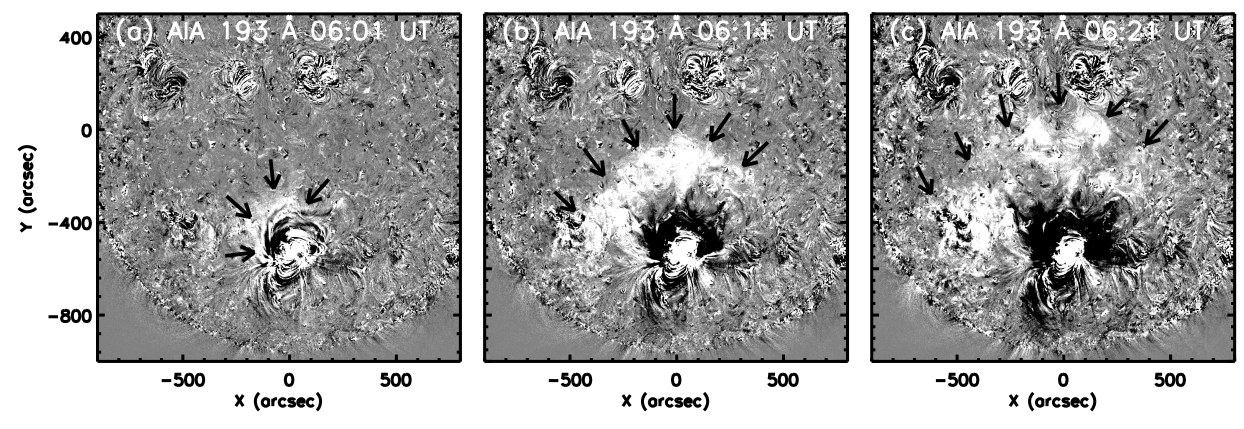}
\caption{Evolution of an EUV wave on 2021 August 28 (a–c) observed in AIA 193 \AA\ base difference images. The base time is 05:45 UT. The black arrows indicate the leading edge of EUV wave front.}
\label{EUV_evolution}
\end{figure*}

As above discussion revealed, there are several questions regarding the physical nature and association of EUV wave including:  What is the association among EUV waves and Moreton waves with other solar activities like CMEs, flares, type II radio burst and SEPs? Why are EUV waves related to ICMEs, geomagnetic storms? and How EUV wave affects space weather? To address these unresolved issues, we aim to analyze these associated phenomena accompanied with EUV waves for Solar cycle (SC) 25 (December 2019 - December 2025).\\ 
%\noindent{\S}{\bf --- Structure of the paper:} 
The paper is organized as follows: Section \ref{sec:obs}  describes the data sources and methodology of EUV waves with associated phenomena. Sections \ref{sec:result} presents the characteristics of EUV waves and Moreton waves, along with their associations with CMEs, solar flares, type II radio bursts, SEPs, and ICMEs. Section \ref{sec:summary} presents the summary and the discussion of results.

%%%%%%%%%%%%%%%%%%%%%%%%%%%%%%%%%%%%%%%%%%

\begin{figure*}[!t]  %%%%% Figure 3
\centering
\includegraphics[width=1\textwidth]{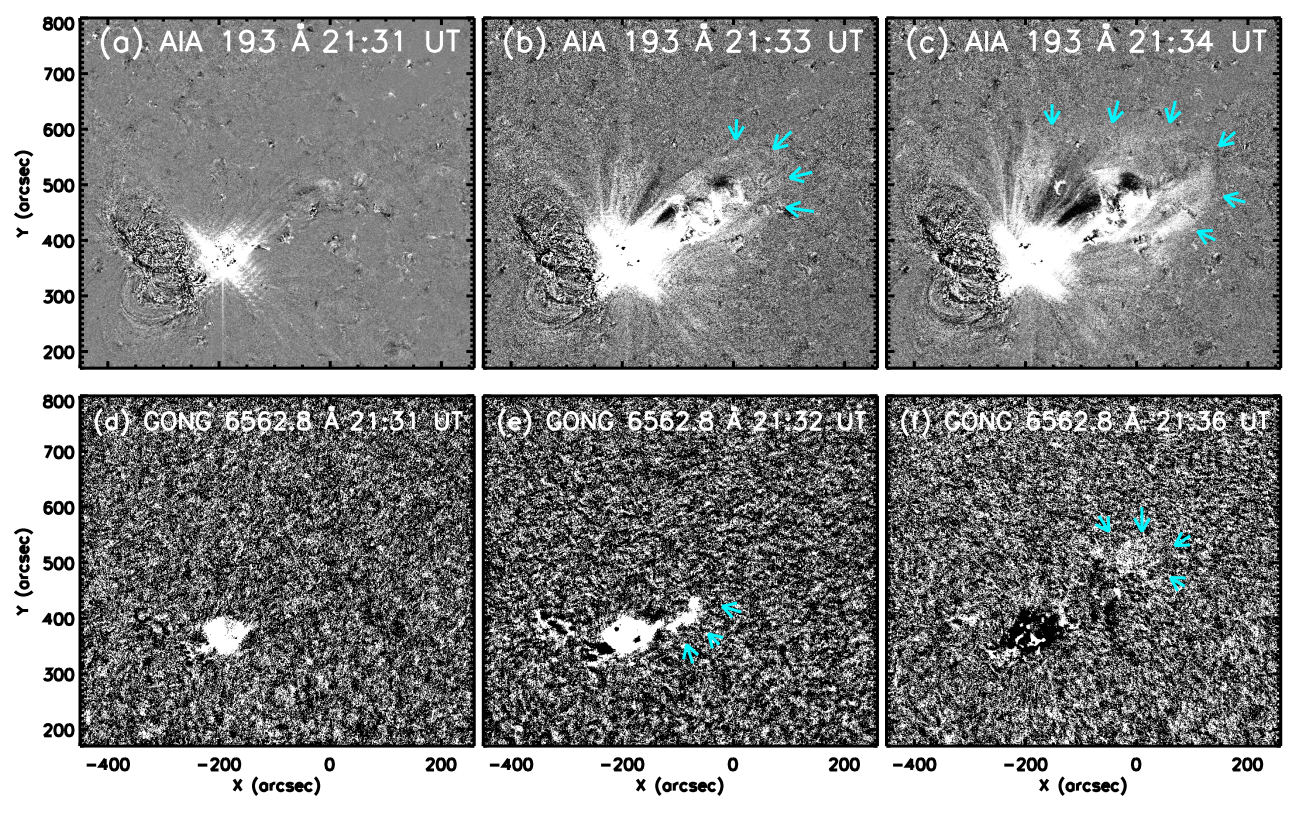}
\caption{The Evolution of an EUV wave in AIA 193 \AA\ (top panel) and associated Moreton wave in H$\alpha$ (bottom panel) on 2021 May 22. The leading fronts of the EUV wave and Moreton wave are shown by cyan arrows.} 
\label{moreton_evolution}
\end{figure*}

\section{Observations and Methodology} 

\label{sec:obs} 
In this study, we investigated a total of 357 EUV wave events observed between December 2019 and December 2025 during the current Solar Cycle 25.
For our analysis, data were obtained from the following sources:
\begin{itemize}

    \item {\bf EUV and Moreton Wave Data:} For the EUV wave data selection during the current solar cycle 25, we have gone through all the this cycle on the catalogue: {\url {https://www.lmsal.com/nitta/movies/AIA_Waves/oindex.html} \citep{Nitta_2013}}. This catalog contains list of EUV events observed by the Atmospheric Imaging Assembly \citep[AIA;][]{Lemen12} onboard the Solar Dynamics Observatory \citep [SDO;][]{Pesnell12} satellite in various UV and EUV wavelengths.  
    %Finally we  have selected the events where the EUV wave is clearly visible. 
For the related Moreton waves, we adopted the list compiled by \cite{Zhong_2025}, which includes all M- and X-class solar flares related Moreton waves observed from December 2019 to December 2023. 
Additionally, for the remaining Moreton waves of B- and C- class from December 2019 to December 2023 and B- to X- class from January 2024 to December 2025, we checked the H$\alpha$ movies of data from the Global Oscillation Network Group \citep[GONG;][]{Harvey1996}.

\item{\bf CME and SEP Data:} The CME data is taken from the Large Angle and Spectrometric Coronagraph CME catalog: \url {https://cdaw.gsfc.nasa.gov/CME_list} \cite{Gopalswamy2009}. The catalog provide the CMEs movies in C2 and C3 coronagraph together with different derived parameters of CMEs such as: onset time, speed, acceleration, angular width etc. The related solar energetic particles (SEPs) are also available in this catalog
EUV waves were associated with CMEs using a time window of 30 minutes to 1.5 hour between the EUV wave onset and the first appearance of the CME in the LASCO/C2 field of view, which begins at about 2 solar radii. To see this association, we have computed the time of the eruption on solar surface using the CME speed provided in the catalogue. Since SEPs are accelerated during solar eruptions, we applied the same time-window criteria as for the flare association i. e. $\pm 15$ minutes.
Proton flux data were obtained from the GOES instruments. We use the integral proton flux in the >10 MeV channel. An event is identified as a SEP event when the peak proton flux in this channel reaches or exceeds 1 pfu (1 particle cm $^{-2} $ s $^{-1} $ sr $^{-1} $) and is at least three times the pre-event background flux.

\item{\bf Flare Data:} To investigate the source region of EUV waves, We have collected the data from several websites. These sites includes: EUV wave catalog, CME catalog as well the information available at Solar Monitor and NOAA solar active region summary reports at NGDC (National Centers for
Environmental Information) site: \url{https://www.ngdc.noaa.gov/stp/space-weather/swpc-products/daily_reports/solar_event_reports/}.
Since EUV waves and flares typically start at nearly the same time, we associated each EUV wave with a flare whose onset time was within $\pm 15$ minutes of the EUV wave onset. We have also analysed all the movies to see the association between EUV wave and solar flares.

\item{\bf Type-II Data:} The related metric (m) type II radio burst data is collected from the catalogue:  \url{https://www.ngdc.noaa.gov/stp/space-weather/swpc-products/daily_reports/solar_event_reports/}.
Since type II radio bursts typically begin after flare onset, and their onset depends on the time and height of shock formation, we adopted a time window of about 1.5 hours for associating them with flares, similar to the approach used for associating EUV waves with CMEs.
% The associated DH type II data is taken from \url{https://cdaw.gsfc.nasa.gov/CME_list/radio/waves_type2.html}. For 2025, DH type II have been identified visually from STEREO-WIND data. 

 \item{\bf ICME Data:} The associated interplanetary coronal mass ejection (ICME) data and their corresponding geomagnetic storm intensity (Dst index) were obtained from the Near-Earth ICME catalog compiled by Richardson and Cane \citep{Richardson_2010}. The catalog is available at: \url{https://izw1.caltech.edu/ACE/ASC/DATA/level3/icmetable2.htm}.
To associate ICMEs with EUV wave events, we examined all ICMEs arriving within 3–5 days after the onset time of each EUV wave. The Richardson-Cane catalogue provides information of associated CMEs with ICMEs.

\end{itemize}

\begin{figure*}[!t]  %%%%% Figure 4
\centering
\includegraphics[width=0.996\textwidth]{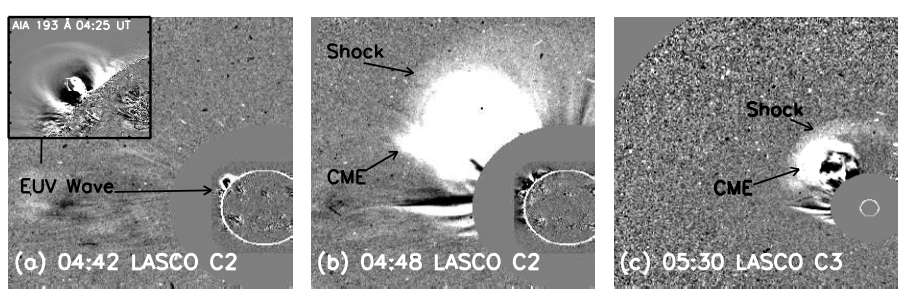}
\caption{An EUV wave event on 2022 January 12 associated with a coronal mass ejection (CME). (a) EUV wave observed in the SDO/AIA 193 Å channel. (b, c) The associated CME observed by SOHO/LASCO C2 and C3, respectively. }
\label{CME_evolution}
\end{figure*}

The association of different Solar activity namely Moreton wave, CMEs, Solar flares, Type II radio bursts, SEPs and ICMEs are described in following subsection. The summary of EUV waves associated activities are presented in Table  \ref{table}. The table contains the number of associated events alongwith their percentage.

\section{Results}
\label{sec:result} 

\subsection{EUV and Moreton Waves} 

  Figure~\ref{Year} presents the year-wise occurrence of EUV and Moreton waves observed between 2019 and 2025.
  The total number of EUV and Moreton waves were 357 and 44, respectively. The histogram shows that EUV waves were significantly more frequent than Moreton waves throughout the study period. Quantitatively, only $\sim$ 12.3\% of the EUV waves were connected to the Moreton waves.
  Very few events were recorded during 2019 and 2020, corresponding to the low solar activity at the beginning of Solar Cycle 25. The number of EUV wave events increased sharply in 2021, reaching its maximum with 28.0 \% of the total recorded events, and then gradually decreased to 19.0\%, 16.2\%, 17.6\%, and 15.3\% in 2022, 2023, 2024, and 2025, respectively.
  
The number of Moreton waves also increase from year 2021 (15.9\%) to 2024 (27.3 \%). Further, the number of events 
  reduced again to 11.3 \% in 2025.
% Similar to EUV wave events, the number of Moreton waves also increase from year 2021 (15.9\%) and 25 \% . In 2024, the percentage of Moreton waves becomes 27.3\% of the their total number. Further, the number of events In 2023 moreton wave are 25 \%
\begin{figure*}[!t]  %%%%% Figure 5
\centering
\includegraphics[width=0.999\textwidth]{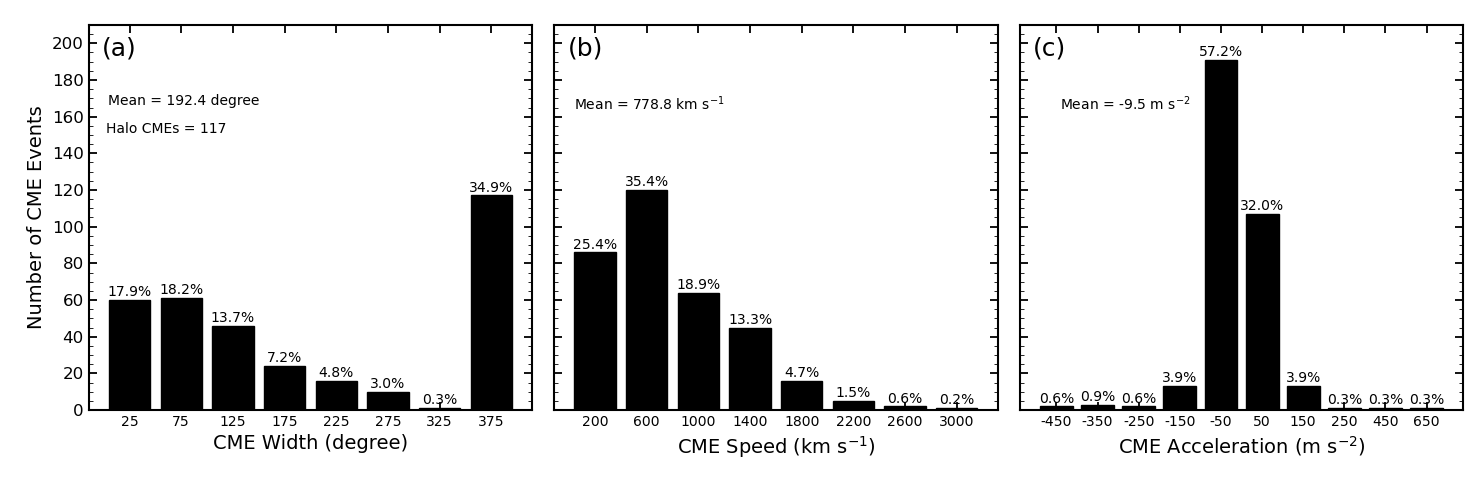}
\caption{Histograms of angular width, speed and acceleration of the CMEs associated with EUV waves.}
\label{CME_distributio}
\end{figure*}

\begin{table*}[t]
\centering
\caption{ Association of the EUV waves with other solar activities 
}
\begin{tabular}{clllc}
\toprule
S.N. & Associated phenomena & Total number & Percentage & Remark \\
\midrule
1 & Moreton Waves           & 44  & 12.3\% &  \\

2 & CMEs                  & 339    & 95\%      &  Halo CMEs 117\\

3 & Solar Flares                  &  331   &   92.7 \%      & \\

4 & Filament/Prominence Eruptions                  & 340    & 95.2 \%       &  \\

5 & SEPs                    & 42  & 11.8\%   & For SEPs $>10$ MeV  \\

6 & m-Type II Radio Burst    & 158 & 44.3\% & \\

7 & DH type II radio burst                  & 116   &   32.5\%      & \\

8 & ICMEs                  & 32    &   9\%      & \\

9 & Geoeffective          & 22  & 6.1\% &  For $|Dst| \geq 50$ nT \\

10 & Active Region Association  & 337 & 94.4 \% & \\

%11 & Clear Visible Shocks    & 24  & 6.7\% & \\
\bottomrule
\end{tabular}
\label{table}
\end{table*}
%CAN BE MOVE TO DISCUSSION SECTION: Overall, the increasing occurrence of both EUV and Moreton waves after 2020 reflects the enhanced solar activity associated with the rising and maximum phases of Solar Cycle 25, with EUV waves occurring much more frequently than Moreton waves.

An example of the evolution of an EUV wave event observed on 28 August 2021 in AIA 193 Å is shown in panels (a)–(c) of Figure \ref{EUV_evolution}. The images are base-difference images obtained by subtracting the pre-event image from each subsequent image, thereby enhancing the visibility of the propagating EUV wave front and its evolution.
% An example of EUV wave  event on 08 August 2021  evolution in AIA 193 \AA\ is dipicted in Panel (a) to (c) of Figure \ref{EUV_evolution}. The images are the base difference obtained by subtracting the pre-event image.
 % Here, we used image difference technique to enhance dynamic wavefronts. In this technique a pre-event image is substracted from subsequent images.
Figure~\ref{moreton_evolution} shows the temporal evolution of the EUV wave and its associated Moreton wave observed on 22 May 2021. Panels (a–c) present base-difference images from SDO/AIA 193 Å, illustrating the outward propagation of the EUV wave from the flare source between 21:31 UT and 21:34 UT. The expanding bright front becomes increasingly prominent with time, indicating the propagation of the coronal disturbance. Panels (d–f) display the corresponding GONG H$\alpha$ line center observations over the same time interval. The arrows mark the EUV wave front and faint Moreton wave front propagating across the chromosphere in association with the EUV wave.

\subsection{CME Characteristics} 
In this section, we studied the CMEs associated with the EUV and Moreton waves. Figure \ref{CME_evolution} depicted one of the examples of the CME and related EUV wave on 12 January 2022. The event was associated with a prominence eruption on the North-East limb. In the figure, we can see the EUV wave ahead of the CME, marked as shock with the arrows.  
Panel (a) of the figure shows the early phase. The inset in the top-left corner is an SDO/AIA 193 \AA\ base-difference image at 04:25 UT, catching a distinct, propagating EUV wave close to the solar limb. The main background image from LASCO C2 shows the very initial emergence of the associated CME into the lower corona.
In panel (b) the CME expands drastically into a classic, bright bubble-like structure. Crucially, a faint, diffuse outer boundary is marked as a Shock, indicating that the CME is moving faster than the local Alfv\'en/sound speed, driving a coronal shock wave.
In panel (c), as the eruption propagates further into the outer corona (observed via the wider field-of-view LASCO C3), both the structured CME core and the leading Shock front remain clearly distinguishable as they expand outwards.

We found 339 out of 357 EUV (about 95 \%)  wave events shows association with CMEs. Moreover, most of the time they are also associated with solar flares. As mentioned in Section \ref{sec:flare_char}, the associated flare class varies from B to X-class.

The distributions of physical parameters of CMEs associated with the sampled EUV wave events are presented in Figure~\ref{CME_distributio}, where all three panels share a common vertical axis representing the  number of CMEs. 
Panel~(a) illustrates the CME angular width distribution, showing a prominent population centered around $25^\circ$ ($17.9\%$) and $75^\circ$ ($18.2\%$) that steadily declines toward $325^\circ$. However, the distribution is heavily dominated by a sharp peak at the final $375^\circ$ bin ($34.9\%$), highlighting a strong statistical association with full or near-full halo CMEs. Panel~(b) outlines the linear speed of the CMEs. The largest fraction of events peaks within the 
$600\text{ km s}^{-1}$ bin ($35.4\%$), with the vast majority of CMEs clustering between few hundred $ \text {km s}^{-1}$ to $1600\text{ km s}^{-1}$, while exceptionally 
high speeds exceeding $2000\text{ km s}^{-1}$ account for less than $3\%$ of the total sample. 
Lastly, Panel~(c) displays the distribution of CMEs acceleration, revealing a highly concentrated spread centered tightly around zero. The overwhelming majority of events fall into the $-50\text{ m s}^{-2}$ ($57.2\%$) and $+50\text{ m s}^{-2}$ ($32.0\%$) bins, indicating that most EUV wave-associated CMEs propagate at nearly constant velocities within the observed field of view.

%speed more and less than 500 km/s. 
\begin{figure*}[!t]  %%%%% Figure 6
\centering
\includegraphics[width=0.95\textwidth]{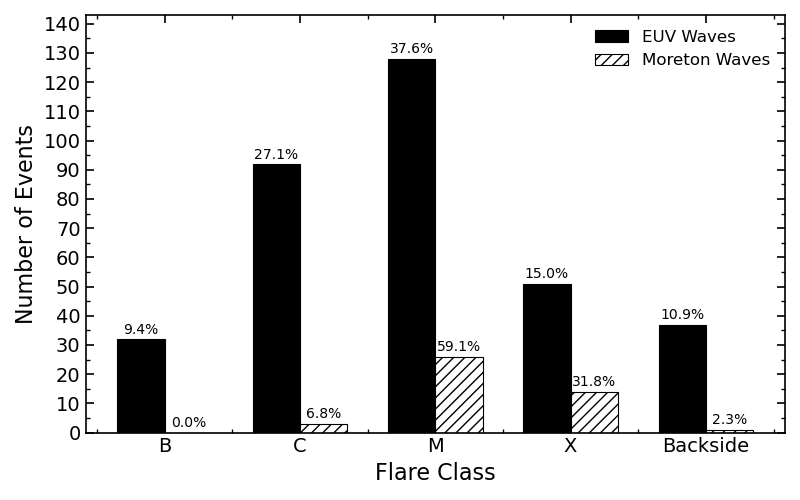}
\caption{Distribution of EUV waves and Moreton waves with  GOES flare class.}
\label{Flareclass}
\end{figure*}

\subsection{Flare Characteristics} 
\label{sec:flare_char}
Figure \ref{Flareclass} illustrates the distribution and percentage of solar flare classes from GOES associated with 340 EUV waves (represented by the black bars) and 44 Moreton waves (represented by the hatched bars). 
{The flare class data is not available for the remaining 17 EUV events.}
The x-axis categorizes the associated solar flares by their GOES classification (B, C, M, and X-class) alongside eruptions originating on the solar backside, while the y-axis tracks the absolute number of observed events. 
 
From the figure, it can be seen that M-class flares emerge as the most frequent triggers, accounting for 128  EUV waves (37.6\%) and a dominant 26 (59.1\%)  Moreton waves. The 32 (9.4\%)  EUV waves are also seen to be triggered by weaker B-class and 92 (27.1\%) by C-class flares, and maintain a steady presence during 51 (15.0\%) high-energy X-class and 37 (10.9\%) backside events.
In contrast, Moreton waves exhibit a clear observational selection effect, showing a strict dependency on high-energy, impulsive eruptions. They are completely absent during weak B-class events and poorly represented in only 3 (6.8\%) C-class flares, with nearly 91\% of all Moreton wave events concentrated exclusively within powerful M- and X-class flares combined.

For backside events, the source location and flare class can not be determined with certainty. Furthermore, the reported flare onset time may not represent the actual onset, as the flare may have begun earlier than the reported time.

\begin{figure*}[!t]  %%%%% Figure 7
\centering
\includegraphics[width=0.999\textwidth]{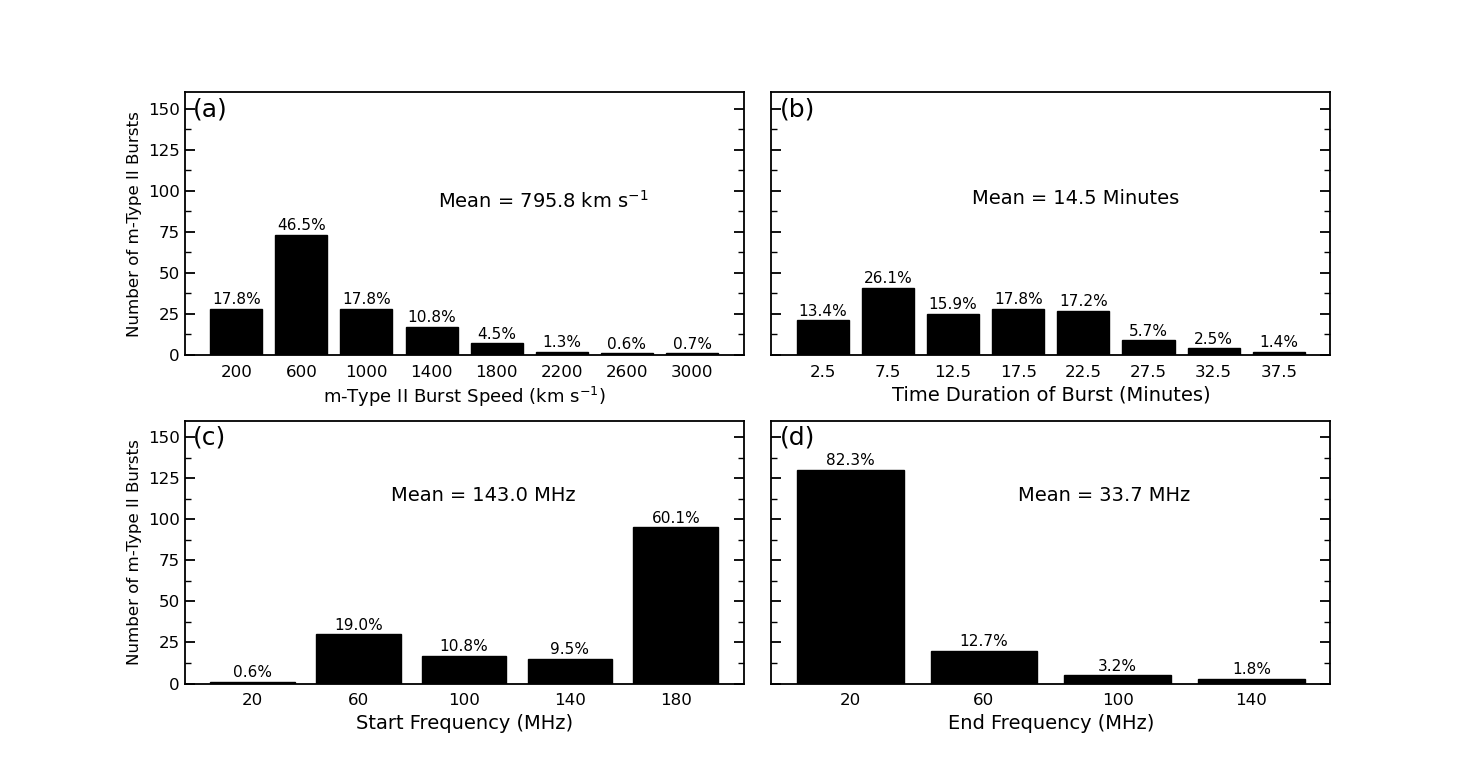}
\caption{Partition of speed, time duration, starting frequency and end frequency of the m-type II radio burst associated.}
\label{TypeII}
\end{figure*}

\subsection{Type II Radio Burst Characteristics} 
The type II radio burst are signature of shock wave expanding through the corona. We found about 44.3\% of the EUV waves are accompanied by m-type II radio bursts (see Figure~\ref{EUV_association}).
We have directly taken the speed of m- type II radio bursts from the active region summary report of NGDC mentioned in section \ref{sec:obs}.
%The reported speeds are computed from the Newkirk density model \cite{Newkirk_61}.}
%For analysis of type II radio burst, we have collected all related parameters from the Radio Solar Telescope Network (RSTN) archival data \url{https://www.ngdc.noaa.gov/stp/space-weather/swpc-products/daily_reports/solar_event_reports/}.type II radio burst speed, starting frequency, end frequency and time duration information from RSTN \url{https://www.ngdc.noaa.gov/stp/space-weather/swpc-products/daily_reports/solar_event_reports/}.
Figure \ref{TypeII} presents the distribution of the number of metric (m)-type II burst events with their speed, time duration, starting frequency, and end frequency. Panel (a) displays the m-type II radio burst speed varies from a few hundred km s ${^{-1}}$ to 3200 $km~s{^{-1}}$, with the majority of events (46.5\%) occurring within the range of 400 km s${^{-1}}$ to 800 km s${^{-1}}$. 
The time durations of m-type II radio bursts are calculated as the difference between end time and start time. Our analysis reveals that all m-type II radio burst associated with EUV waves exhibits maximum drift duration of upto 40 minutes as shown in Panel (b). Most radio bursts (26.1 \%) have time differences between 5 to 10 minutes in their start and end frequency. Panel (c) and (d) show start and end frequency plots of m-type II radio burst, respectively. 
The RSTN observed the radio bursts in the frequency range of 25-180 MHz, so these are the lower and upper bounds of the frequency range in our sample.

Maximum events show highest start frequency of 180 MHz and least end frequency of 25 MHz. Here, 72 radio bursts span the entire range of radio-spectrograph (25-180 MHz) from RSTN.  We also examined the continuation of m-type II radio bursts into DH-type radio bursts.%finding a rate of approximately 58\%. 
Therefore, the association between EUV waves and the DH type II radio bursts becomes 32.5\%.

\begin{figure*}[!t]  %%%%% Figure 8
\centering
\includegraphics[width=0.999\textwidth]{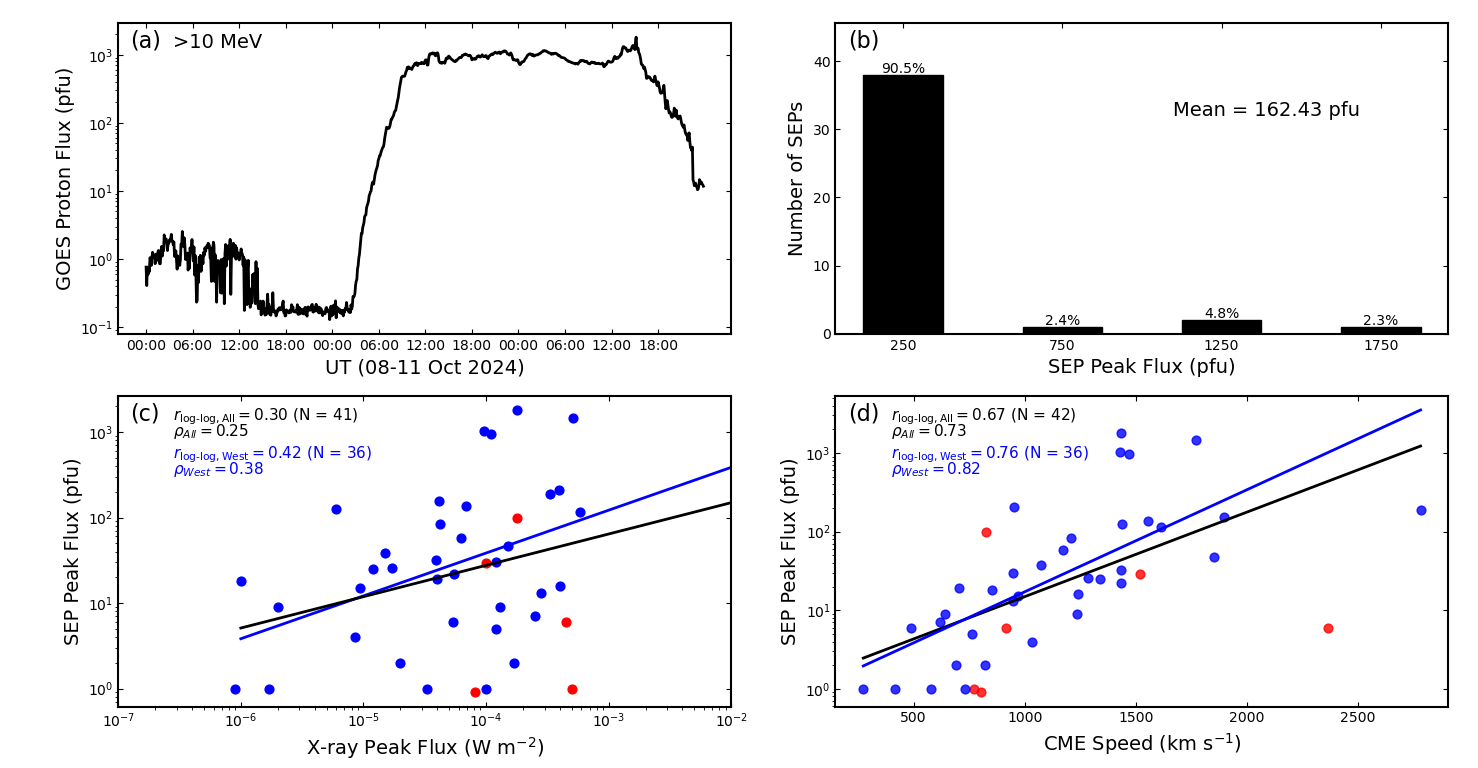}
\caption { Panel (a): An example of high intensity SEP event of 09 October 2024., Panel (b): Histogram of Number of SEPs with peak proton flux (pfu).
Panel (c), (d): Scattered plot of SEPs peak proton flux with X-ray peak flux and CME speed. The values of log-log Pearson correlation coefficients for all events ($r_{\mathrm{log-log, All}}$) and for western events ($r_{\mathrm{log-log, West}}$) are shown with sample size (N) at the top left of each panel. Here, $\rho_{\mathrm{ All}}$ and $\rho_{\mathrm{ West}}$  represents the Spearman correlation coefficient calculated for all SEPs events and for western events respectively. The blue and red data points indicate events originating from the eastern and western hemispheres, respectively. The overlaid black and blue line show linear fit of all data points and blue points, respectively.  }
\label{sep}
\end{figure*}

\subsection{SEPs Characteristics} 

The SEPs are driven either by shocks propagating ahead of CMEs or by magnetic reconnection processes during solar flares. Therefore, it is worth to see how the SEP peak flux is related to the %we analyzed association between SEP Peak flux, 
GOES X-ray flux and CME speed. 

The statistical characteristics and a representative case study of SEP events are compiled across four subplots in 
Figure~\ref{sep}. 
Panel (a) presents a temporal variation of the proton flux from GOES for energy $>$ 10 MeV, showing a SEP event during 08-11 October 2024. The plot 
%Panel~(a) presents one example of the time-series evolution of the $>10\text{ MeV}$ GOES proton flux over the period from October 8 to October 11, 2024, which
captures a quiet baseline phase followed by an abrupt increase % multi-order-of-magnitude surge 
early on October 9 that peaks at 1810 pfu before undergoing a slow decay.
Panel~(b) presents the distribution of peak proton flux of the SEPs associated with EUV waves ( 11.8 \%). %histogram categorized by SEP peak flux ranges in proton flux units (pfu), 
The histograms shows
%demonstrating 
that lower-intensity events dominate the sample; specifically, $90.5\%$ of events fall within the 
$0\text{--}500\text{ pfu}$ range, $2.4\%$ within $500\text{--}1000\text{ pfu}$, and $4.8\%$ within $1000\text{--}1500\text{ pfu}$, which again drops to 2.3 \% in range of 1500 -- 2000 pfu. The mean flux of all these SEPs is found to be 162.4 pfu.
%whereas higher-intensity intervals display a sharp drop-off before a minor rebound to $2.5\%$ in the $1500\text{--}2000\text{ pfu}$ bin with average of SEPs peak flux of 166 pfu as displayed in panel .  

Next, we check how the proton flux is related to the X-ray flare flux and CME speed. 
To do so, we compute the log-log Pearson correlation coefficient  of proton flux with the latter two. The distribution is displayed as scattered plots in panels~(c) and (d) respectively. Panel~(c) shows the log SEP peak flux against the log flare X-ray peak flux ($\mathrm{W\,m^{-2}}$). For all events ($N = 41$), the log-log Pearson correlation ($r_{\log\text{-}\log,\mathrm{All}}$) between
X-ray peak flux and SEP peak flux is 0.30 (p value = 0.05, 95\% confidence interval (CI): $-0.007$ to 0.557), and the Spearman rank correlation ($\rho_{\mathrm{All}}$) is 0.25 (p value= 0.11, 95\% CI: $-0.068$ to 0.528).
However, both correlation coefficients are not statistically significant at the 5\% level. For west-limb events ($N = 36$), $r_{\log\text{-}\log,\mathrm{West}}$ is 0.42 (p value = 0.01, 95\% CI: 0.107 to 0.658) and $\rho_{\mathrm{West}}$ is 0.38 (p value = 0.024, 95\% CI: 0.044 to 0.636). Both are statistically significant at the 5\% level, although the CIs are wide and the association is only moderate. Here, N and CIs are sample size and Confidence Interval at 95 \%, respectively. The GOES X-ray flare class and the corresponding peak flux are not available for one event. 

Panel~(d) shows the SEP peak flux against the CME speed. 
The Pearson correlation between CME speed and log SEP peak flux is 0.61 (p value < 0.001, 95\% CI:
0.376 to 0.771) for all events (N = 42)  and 0.72 (p value < 0.001, 95\% CI:
0.511 to 0.847) for west-limb events (N = 36). For all events ($N = 42$), $r_{\log\text{-}\log,\mathrm{All}}$ between CME speed and
SEP peak flux is 0.67 (p value < 0.001, 95\% CI: 0.459 to 0.809) and
$\rho_{\mathrm{All}}$ is 0.73 (p value < 0.001, 95\% CI: 0.528 to 0.851),
indicating a strong positive association. For west-limb events ($N = 36$),
$r_{\log\text{-}\log,\mathrm{West}}$ is 0.76 (p value < 0.001, 95\% CI: 0.570 to
0.869) and $\rho_{\mathrm{West}}$ is 0.82 (p value < 0.001, 95\% CI: 0.658 to
0.893). Both coefficients indicate a strong, statistically significant
positive association. The $p$-values are calculated using a Student's $t$ distribution with $N-2$ degrees of freedom, and statistical significance was assessed at the 5\% level. The 95\% confidence intervals (CIs) for Pearson's coefficients were computed with the Fisher $z$-transformation. The 95\% CIs for Spearman's $\rho$ were estimated using a percentile bootstrap method, based on 10,000 resamples of the paired observations drawn with replacement.
 The correlation coefficients and sample sizes given above and in Figure \ref{sep} are consistent with the event data listed in Table \ref{table2} given in Appendix.

\begin{figure*}[!t]  %%%%% Figure 9
\centering
\includegraphics[width=0.999\textwidth]{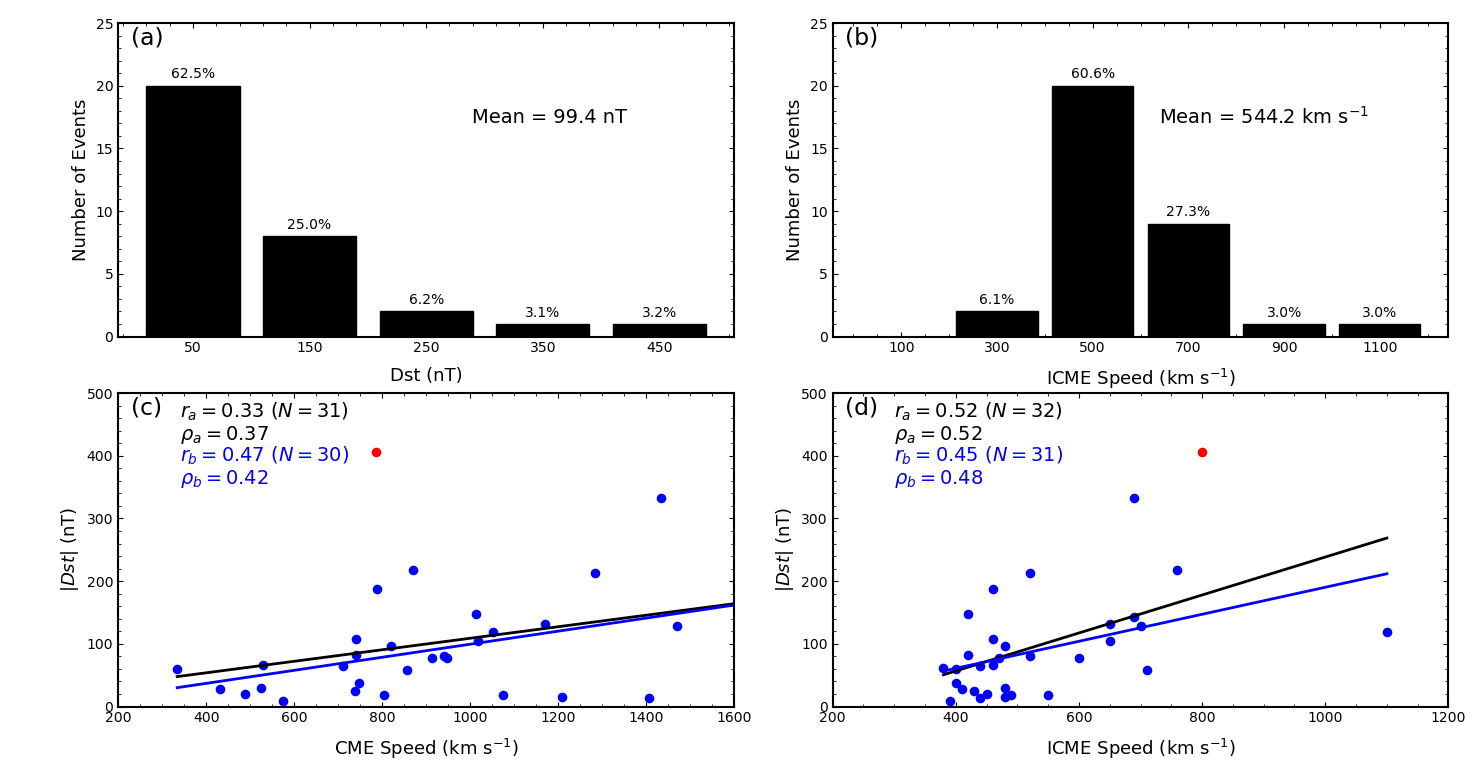}
\caption{ Panel (a), (b): Histogram showing the distribution of ICMEs events with the $ |Dst|$ index (nT) and ICME speed, respectively. Panel (c), (d): Scatter plots of $ |Dst|$  (nT) versus the  CME speed (km s$^{-1}$) , and ICME speed (km s$^{-1}$), respectively. 
 The linear Pearson correlation coefficients are represented as $r_{\mathrm{a}}$  (for all events ) and 
$r_{\mathrm{b}}$  (for less than 400 nT events) with sample size (N) at the top left of each panel. Here, $\rho_{\mathrm{a}}$ and $\rho_{\mathrm{b}}$ show the Spearman correlation coefficient calculated for all  events and for less than 400 nT events respectively.
The overlaid black and blue line represent linear fit of all data points and blue points respectively. The red dot and blue dot represent $|\mathrm{Dst}| \geq 400$~nT events and less than 400 nT events respectively.}

\label{ICME}
\end{figure*}

The SEPs events are divided in two groups to calculate correlation between different SEPs characteristics. Firstly, all events are taken in account, and second, events that originated in western solar hemispehere. %The values of correlation coefficient (r) is shown at the top left of each panel. 
When we see the correlation of these quantities taking into account the events from the west hemisphere, the values of correlation coefficient increases.

\subsection{ICMEs and DST Characteristics}
To understand the impact of EUV waves on space weather, we examined their association with interplanetary coronal mass ejections (ICMEs) and the corresponding Dst index as an indicator of geoeffectiveness. We identified 32 (9\%) EUV wave events associated with ICMEs.
Among these, 13 ICMEs were associated with Moreton waves, accounting for 29.5\% of the total observed Moreton waves in our list.
The distributions of observed ICMEs and geoeffectiveness are presented in Figure~\ref{ICME} `a' and `b' panels. Panel (a) shows the distribution of ICMEs events categorized by the magnitude of the Dst index. 
More than 37.5\% of the ICME events  (12 out of 32) are associated with strong geomagnetic storms ($|Dst| \geq 100$ nT).

Panel (b) represents the distribution of ICMEs speed with number of events. The range of speed of ICMEs varies upto $1200\text{ km s}^{-1}$ with majority of events (60.6\%) in range 400--600 km s$^{-1}$ and the mean speed of $544\text{ km s}^{-1}$.
%with average speed of 
The relationships between the absolute Dst index with the CME and ICME speed are illustrated in panels (c) and (d), respectively. Out of 32 ICMEs, 31 are associated with geomagnetic storm. The data of one event is missing in the Richardson catalog, so we exclude it from the correlation analysis. %of Dst index with CME speed and ICME speed, as demonstrated in Panel (c) and Panel (d) respectively.
One of the ICME events shows very strong Geomagnetic storm ($|Dst| \geq 400$ nT), shown with red dot, all other ICMEs event with blue dot. 
The linear Pearson correlation ($r_{\mathrm{a}}$) is 0.33 (p value = 0.071, 95\% confidence interval [CI]: $-0.029$ to 0.612). The Spearman rank correlation ($\rho_{\mathrm{a}}$) is 0.37 (p value = 0.038, 95\% CI: $-0.024$ to 0.688), which is marginally significant. After excluding events with $|\mathrm{Dst}| \geq 400$~nT ($N = 30$), the correlations increase to $r_{\mathrm{b}} = 0.47$ ( p value = 0.009, 95\% CI: 0.131 to 0.709) and $\rho_{\mathrm{b}} = 0.42$ (p value= 0.022, 95\% CI: 0.014 to 0.736). The CIs are wide, and the association is moderate.
Panel~(d) shows $|\mathrm{Dst}|$ against ICME speed. For all events ($N = 32$), $r_{\mathrm{a}} = 0.52$ (p value = 0.002, 95\% CI: 0.211 to 0.736) and $\rho_{\mathrm{a}} = 0.52$ ($p = 0.002$, 95\% CI: 0.209 to 0.751). For events with $|\mathrm{Dst}| < 400$~nT ($N = 31$), $r_{\mathrm{b}} = 0.45$ (p value = 0.012, 95\% CI: 0.110 to 0.692) and $\rho_{\mathrm{b}} = 0.48$ (p value = 0.007, 95\% CI: 0.148 to 0.722). All coefficients are statistically significant at the 5\% level and indicate a moderate positive association, although the CIs are wide.
 We exclude extreme events having $|Dst| \geq 400$ nT, because they can result from complex conditions, such as CME-CME interactions, which make it difficult to attribute the geomagnetic response to a single solar source.
All ICME parameters used in this analysis were obtained from the Richardson ICME catalog. 
%The $|Dst|$ index exhibits a strong positive correlation ($r = 0.69$) with the mean magnetic field strength, a moderate correlation ($r = 0.47$) with the ICME speed, and a comparatively weaker correlation with the CME speed. These results suggest that the magnetic field strength of an ICME plays a more significant role in determining its geoeffectiveness than its propagation speed. 

\begin{figure}[!t]  %%%%% Figure 10
\centering
\includegraphics[width=0.994\textwidth]{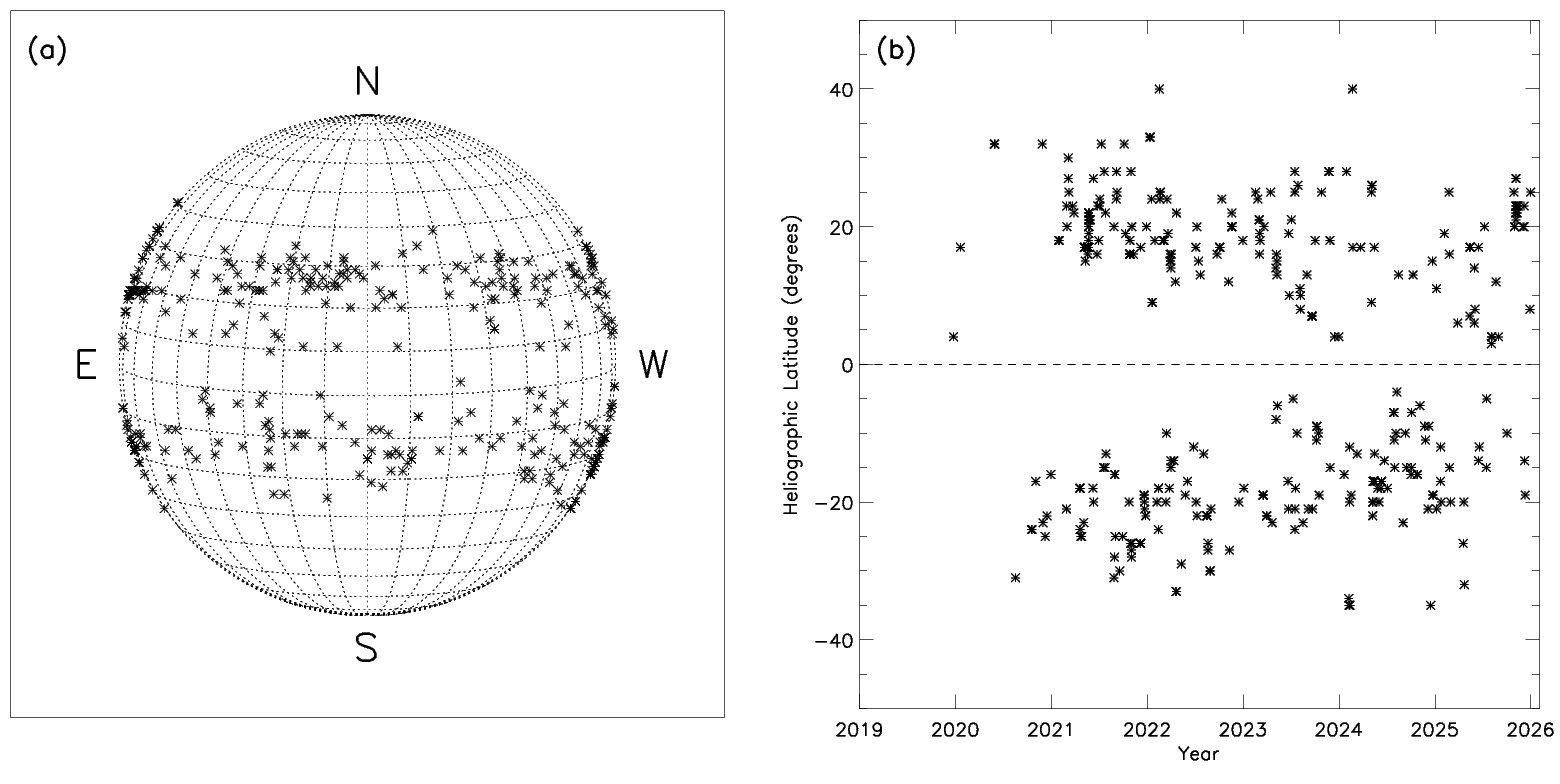}
\caption{Panel (a): Locations of EUV waves sources shown on the solar disk. Panel (b): The latitudinal distribution of EUV waves source locations with time.}
\label{location}
\end{figure}

\subsection{EUV Waves Source Locations}

The source locations of the EUV wave events are identified using the AIA images/movies, and the position of each source region is plotted in Figure~\ref{location}.
Panel (a) shows distribution of EUV waves source locations on solar disk.
We observe 190 (53.2\%) of the events in the Northern Hemisphere and 167 (46.8\%) in the Southern Hemisphere. The source locations are concentrated between $0^\circ\text{--}40^\circ$ in the Northern Hemisphere and $0^\circ\text{--}30^\circ$ in the Southern Hemisphere. These bands correspond to the classic solar activity belts, where sunspots, solar flares, and coronal mass ejections typically originate. 
There is a distinct lack of source regions directly along the equator ($0^\circ$ latitude) and a complete absence of points in the high-latitude polar regions (above $40^\circ$\,N or $40^\circ$\,S). Regarding the longitudinal distribution, the source regions are spread across all longitudinal zones from east to west. 
Panel (b) represents latitude-wise distribution of EUV wave events source region with time of current Solar Cycle. In starting of Solar cycle from 2019 to 2021, occurrence of events are minimum and spreads in higher latitudes dominantly. 
This distribution becomes narrow near to equator of solar disk as cycle progress. This distribution is known as "butterfly diagram".

%We analyzed 357 EUV wave events occurred during SC 25 (from Dec 2019 to Dec 2025). Our analysis primarily investigate association of EUV waves and Moreton waves with CMEs, Solar Flares, m-type II radio burst and SEPs for SC 25 (December 2019- December 2025). 

%The locations of all EUV wave events identified during this period are shown in Figure \ref{location}.

%Most events  originated from northen hemisphere (174 events). Among the 356 events, 28 events originated from backside of disk so their exact locations are not identified. 
%farside of the sun.

%Figure \ref{EUV_association}
%summarizes association of EUV waves and moreton waves with other solar activities ( CMEs, solar flares, m-type II radio burst and SEPs)  in detail. We found that 82.5 \% EUV waves and 91.5 \% Moreton waves are associated with both CME and Solar flares. In addition, 12.8 \% EUV waves are only associated with solar flares, whereas fewer than 2\%\ are merely with CMEs. However, no Moreton waves was detected in absence of flare. m-Type II radio burst and SEPs shows association of 44.6 \% and 12 \% respectively. 

\begin{figure*}[!t]  %%%%% Figure 11
\centering
\includegraphics[width=0.999\textwidth]{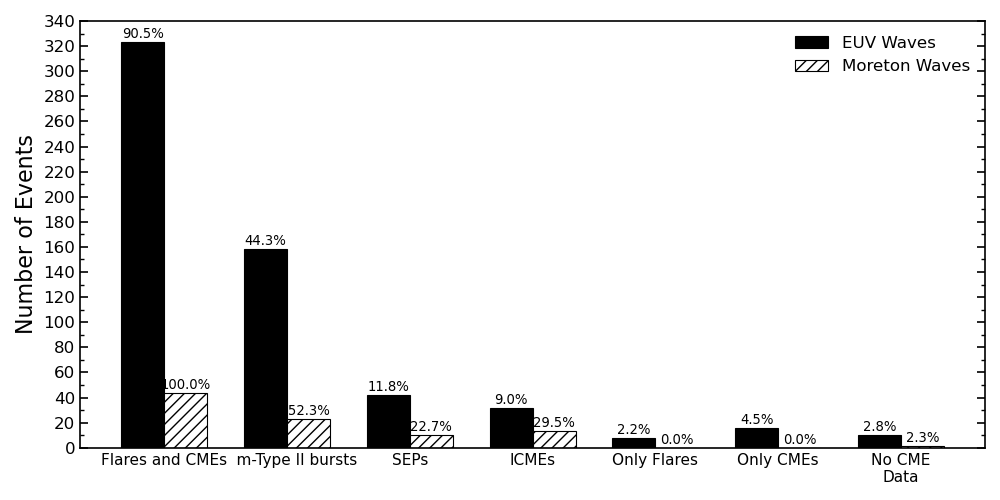}
\caption{Association between EUV waves and Moreton waves with other solar activities.}
\label{EUV_association}
\end{figure*}

\section{Summary and Discussion}
\label{sec:summary} 

In this study, we presented 357 EUV wave events during Solar Cycle 25 and examined their association with other solar activity, namely Moreton waves, CMEs, solar flares, type II radio bursts, SEPs, and geomagnetic storms (See Table \ref{table2} in Appendix). Our main results are summarized as follows:
 
\begin{itemize}
    \item {We identified 357 EUV waves, 44 of which are associated with Moreton waves. }

  \item { The majority of EUV waves are associated with both CMEs and solar flares (90.5\%). 9\% of EUV waves are connected with ICMEs at the Lagrange point L1, and 6.1\% produced geomagnetic storms with $|Dst| \geq 50$ nT. Majority of observed Moreton waves are associated with CMEs and flares greater than M-class.}
 
  \item { 44.3\% of EUV waves are associated with m-type II radio bursts and 32.5\% are associated with DH type II radio bursts. }

 \item {We found only 11.8 \%  EUV wave events are associated with SEP events. Their peak flux intensity ranges from few to 2000 pfu.}

\end{itemize}

The association between EUV waves and Moreton waves is crucial to understanding their underlying physical phenomena. Figure \ref{EUV_association} presents distribution of solar activities associated with EUV waves (shown by black bars) and Moreton waves (shown by hatched bars). Total 323 EUV events (90.5 \%) EUV and 43 Moreton events (98 \%)  are associated with both CMEs and flares. The occurrence decreases for type II radio burst, SEPs and ICMEs. Eight EUV waves (2.2 \%) occur with flares only without CMEs, whereas 16 EUV waves (4.5 \%) are accompanied with CMEs without flares. The CME data are not available for 10 EUV waves (2.8 \%) and one Moreton wave.  

Case studies on the connection between EUV waves and Moreton waves have been performed in the past \cite{Asai12,Seki2021,Cabezas19,Warmuth2004,Long_2019,Zhong_2025,Zheng_2023}.
\citep{Asai12} studied EUV and Moreton waves associated with a GOES X6.9-class solar flare. They found that the fast component of the EUV wave was cospatial with the Moreton wave. They also found that the wave was associated with a type II radio burst, and that the speed of the type II radio burst was comparable to that of the EUV wave's fast component.
\citep{Warmuth2004} analyzed 12 Moreton waves; EUV data from EIT onboard SOHO were available for 8 of these events, and all were found to be associated with EUV waves.
\cite{Long_2019} reported four EUV waves from AR 12017 associated with flares of GOES M-class or greater. All of these EUV waves were associated with Moreton waves. \cite{Zhong_2025} studied 66 Moreton waves, all of which were associated with EUV waves.
\cite{Zheng_2023} studied eight EUV waves associated with Moreton waves. They explained that this association was stronger when the filament eruptions were highly inclined, with the nose of the fast-component EUV wave close to the solar surface.
In our present study, the 12.3\% association between EUV waves and Moreton waves supports the earlier, small statistical samples indicating that only a small fraction of EUV waves are accompanied by Moreton waves. In contrast, all Moreton waves are necessarily associated with EUV waves. 
Moreton waves are the chromospheric counterparts of the coronal fast-component EUV waves. When the fast-component EUV waves propagate out in the solar corona, they also travel down to the chromosphere as a bubble. However, we do not see Moreton waves in every coronal EUV wave event because the chromosphere is much denser than the corona. Due to the high density of the chromosphere, not all the fast-component EUV waves manage to penetrate through the chromospheric plasma \citep{Chen_2016, Zhong_2025}. This is the reason we find less association between these two.

In a statistical analysis during 1997--1998 \cite{Biesecker2002} found all the EIT waves observed by EIT/SOHO telescope were accompanied by CMEs. In case of solar flares this association is not very strong. In case of \cite{Muhr2014} 95\% of EUV waves  were associated with CMEs and  74 \% are associated with flares.
\cite{Nitta2014} studied 34 EUV waves from the EUVI data between March 2007 and December 2009 out of which twenty one (62 \%) were associated with CMEs. They showed that almost all the EUV waves were connected with solar flares of GOES class A to X class. If we remove the GOES A and B class flares, their EUV wave and flare association becomes 41.1 $\%$. 
According to the study of \cite{Chen2006} CMEs are more important to generate EUV waves rather than strong (M and X class) compact flares. 

In our study, 339 (95 \% ) EUV waves are accompanied by CMEs. In the remaining 18 events (5 \%), 10 have no LASCO CME data and 8 events have no association with CMEs. If we exclude the EUV wave events without the coronagraph data, the association rate becomes 97.7 $\%$. So in summary, we can say that majority of EUV wave are associated with CMEs, which confirm the finding of \cite{Biesecker2002,Muhr2014}. 
 A total of 340 EUV waves (95.2 \%) are associated with solar flares, among which 331 flares appeared on the solar disk as listed in Table \ref{table}, and another nine events were on the backside.
%{\bf Total 331 EUV waves (92.7 \%) are associated with solar flares, as listed in Table \ref{table}. The association was determined by inspecting the movies available online from the catalogue, as described in section \ref{sec:obs}. In section \ref{sec:flare_char}, we mentioned a total of 340 EUV waves associated with flares. The difference is because of the fact that the source regions for 09 events are extreme backside, so we could not see the flare.}
The flare class varies from GOES B to X classes. If we exclude small B-class flares from our dataset this percentage becomes 83.8\%.
In case of Moreton wave, 43 (98 \%) are accompanied by CMEs and all are accompanied by solar flares. For one associated Moreton wave, CME data are unavailable.
As far as the flare association, the Moreton waves are associated from GOES C to X class. However, the majority are associated with $>M1.0$ GOES class flares confirming the earlier findings \citep{Asai12,Long_2019,Warmuth2004,Zhong_2025}. 

As the fast-component EUV wave and type II radio burst are the signature of the same fast mode MHD shock considered as fast mode shock, a strong correlation in their occurrence is expected. \cite{Klassen2000} collected 21 type II radio bursts of 1997, then EIT data corresponding to the time interval analyzed. 
They reported 19/21 (90 \%) type II radio burst were associated with slow-component EUV waves. \cite{Muhr2014} analyzed 60 EUV waves (January 2007 to February 2011) and found 11/49 (22.4 \%) for coronal and 9/60 (15 \%) for interplanetary type II bursts. \citep{Nitta_2013} analyzed 171 EUV waves, which they name Large-scale Coronal Propagating Fronts (LCPF) between April 2010 and January 2013 January, 138 arc-shaped fronts propagate across the disk. They found that 75/138 (54\%) of the events were associated with Type II radio bursts. \cite{Long_2017} analyzed 164 EUV waves, and found only 66/164  (40 \%) were associated with type II radio burst.
Very recently \cite{Chandra2026} analysed the correlation between m-type II radio bursts and fast-component EUV wave speed and they found very strong correlation (0.77) between them. 

In our analysis, 158/357 (44.3\%) of EUV waves are associated with m-type II radio bursts.  In the case of DH type II bursts, this association rate is 116/357 (32.5\%).  Here, Moreton waves show a stronger relationship with m-type II radio bursts, with an association rate of 23/44 (52.3\%).
This association rate matches previous studies \citep{Muhr2014,Nitta_2013,Long17}.

From this analysis, we can conclude that when we compare type II radio bursts with EUV/Moreton waves, we find a strong association; however, when we look in the opposite direction (EUV waves and their association with type II radio bursts), the association becomes weaker.
The lower association implies the variety in the shock strength from too weak or no shock for type II burst to strong enough for it.
This association can be further examined by combining different viewpoint observations from Solar Orbiter, which we plan to pursue in our future study.

The physical relationship between EUV waves and SEP events is still an open question. In study of 164 EUV waves by \citep{Long_2017}, only 21/164 (12.8 \%) showed evidence of associated proton event. \cite{miteva2014} analyzed 179 SEP events associated with flares (GOES class $ \geq C9.0$)  during Solar Cycle 23, and reported only fewer (10\%) EIT waves are associated with SEP events, while contrary to this, major SEP events 133/153 (87\%) were associated with EUV waves. In our case the association is only 42/357 (11.8\%).
Our results are consistent with the study of \cite{miteva2014,Long_2017}.
SEPs are stronger when their source region lies toward the western hemisphere, owing to the Parker spiral geometry of the interplanetary magnetic field \cite{Gopalswamy_2014,Long_2017}. The EUV waves in our sample are distributed over the whole solar surface, which is one reason for their low association with SEPs. When we restricted the sample to EUV waves located near the western limb ($>60^\circ$ longitude), the association rate increased to 26\%. A second possible reason is that weak EUV waves are likely poor shocks, and so are inefficient in accelerating particles.

We found lower log-log Pearson correlation of 0.30 between X- ray Peak flux and SEP Peak proton intensity. This result agrees with previous studies, which also reported weak relationship between X-ray flux and SEP proton intensity \cite{Gopalswamy2003_SEP, Gopalswamy2004_SEP,Chandra2013_SEP}. These studies also reported moderate association of proton intensity with CME speed $\sim$ 0.45 \cite{Gopalswamy2004_SEP}, $\sim$ 0.42 \citep{Chandra2013_SEP}. We obtained  Pearson correlation of 0.61 (log SEP flux - CME speed) and 0.67 (log SEP flux-log CME speed) for all SEP events. For SEP events originating from the western solar disk beyond W00 ($>0^\circ$ longitude), the correlation increases to 0.72 (log SEP flux - CME speed) and 0.76 (log SEP flux-log CME speed).

EUV waves are associated with CMEs (as mentioned above) whose interplanetary counterparts (ICMEs)  drive geomagnetic storms.
%\cite{Syed2019} presented statistical analysis of CMEs (near sun region) and ICMEs near 1 AU for SC 23 and SC24 (1996-2017). For this they compare CMEs characteristics with two subsets of ICMEs: Magnetic clouds and Ejecta.
%\cite{Samuel_2023} investigated correlation between various parameters for SC 23 and SC 24 (1996-2019). For this, they analyzed 111 GSs having DST index $|Dst|  \geq 100$ nT). They obtained correlation of 0.30 for CME speed with $|Dst|$ index and 0.62 for ICME speed with $|Dst|$ index. 
%For, $|Dst| $ index  $\geq 100$ nT), \cite{MitewaS_2023} identified 550 events in same time duration and obtained correlation of 0.17 ± 0.07 (between $|Dst|$ index and CME speed) and 0.45 ± 0.07 (between $|Dst|$ index and CME speed).
\cite{Mitewa2023} found no clear association between DST index and CME speed and the correlation coefficient ranges between 0.04 (with LASCO CME Speed) to 0.55 ( with the Graduated Cylindrical Shell model - an approach that derives the geometric and kinematic properties by three-dimensional reconstruction of a CME). For the relationship between ICME speed and DST index, they obtained a correlation of 0.37 and 0.45 using Wind and ACE instruments data respectively. We yield the Pearson correlation coefficient of 0.33 (for all 32 ICME events) and 0.47 (excluding one event having $|Dst|  \geq 400$ nT) between DST index and CME speed . Similarly for DST index and ICME speed, the linear Pearson correlation coefficient is 0.52 (for all events) and 0.45 (excluding one event having $|Dst|  \geq 400$ nT). 
A total of 32 ICMEs registered by the L1 satellites are associated with 46 EUV waves. The difference  is because ICMEs are associated with more than one EUV wave, and may be due to their origin from the solar backside.
%These results are compatible with previous studies. \citep{Samuel_2023,MitewaS_2023,Mitewa2023}
For the weak association of EUV waves with ICMEs and geomagnetic storms in our analysis, many eruptions are not Earth-directed, so we cannot detect the resulting ICME or its geoeffective consequences. Another reason for the weak association is that most CMEs decelerate after reaching a certain height, and only accelerating CMEs can propagate to large heliocentric distances. Furthermore, in the absence of an ICME, there is no or very less chance of getting a geo-magnetic storm.

There are many selection effects which may affect the obtained results. First, Moreton wave identification depends on available H$\alpha$ data, meaning faint signatures can be missed without high-cadence, multi-observatory coverage. Second, the CME parameters obtained from coronagraph observations are subject to projection effects. The observed CME angular width and propagation characteristics can differ from their actual values depending on the CME source location and propagation direction. The SEP detection is affected by the magnetic connectivity between the CME/shock source region and the observing spacecraft. A  SEP event may not be detected if the relevant magnetic field lines are poorly connected to the acceleration region. Finally, the ICME sample is subject to an Earth-directed selection effect because ICMEs are identified in situ near Earth. As the consequence, our ICME sample represents CMEs whose propagation and magnetic structure allow them to encounter the Earth.

Overall, our study highlights the significant role of EUV waves in the occurrence of other  solar eruptive phenomena such as CMEs, solar flares, SEPs, and metric type II radio bursts. 
Although this analysis is limited to EUV wave events observed during the first half of Solar Cycle 25 (December 2019-December 2025), the observed association with these eruptive phenomena indicates that EUV waves may be a useful indicator of space weather impact.

\label{future}

\section*{Author Contributions:}
A.D. did the data analysis and wrote a substantial part of the
manuscript. R.C. and P. D. wrote a significant part of the manuscript and
contributed to the interpretation. 
All authors have read and agreed to the published version of the manuscript.

\section*{Funding:} This research was funded from the ANRF under project number EEQ/2023/000214. P.D. acknowledges the support from the Research Council of Norway through its Centers of Excellence scheme, project number 262622.

\section*{Data Availability Statement:} The EUV events  analyzed in this study are available on {\url {https://www.lmsal.com/nitta/movies/AIA_Waves/oindex.html}} with movies in UV and EUV wavelengths.The LASCO CME images from C2 and C3 and  CME details are taken from \url{https://cdaw.gsfc.nasa.gov/CME_list/}.The type II radio burst informations are available at \url{https://www.ngdc.noaa.gov/stp/space-weather/swpc-products/daily_reports/solar_event_reports/}. The  catalog  used for ICMEs and DST information are compiled by Ian Richardson and Hilary Cane, available on \url{https://izw1.caltech.edu/ACE/ASC/DATA/level3/icmetable2.htm}.
The EUV wave images from AIA files used in this article, are publicly available from the JSOC at \url{jsoc.stanford.edu}. The GONG images in $H \alpha$ wavelength are taken from \url{https://gong2.nso.edu/archive/patch.pl?menutype=hAlpha#step2}. 
% The DH type II burst catalogue is available on \url{https://cdaw.gsfc.nasa.gov/CME_list/radio/waves_type2.html}. 

\section*{Acknowledgment} 
We thank the reviewers for their valuable comments and suggestions which improved the manuscript quality. We acknowledge the open data policy of the  of NASA’s Solar Dynamics Observatory mission, Radio Solar Telescope Network, Global Oscillation Network Group and e-Callisto
data. We also acknowledge the LASCO-SOHO providing the archival data.
%A.D. and R.C. acknowledge support from the DST/SERB project number EEQ/2023/000214

\section*{Conflicts of Interest:} The authors declare no conflicts of interest.

\vspace{-1em}

\PublishersNote{}
\isPreprints{}
%{% This command is only used for ``preprints''.
%\end{adjustwidth}
%} %If the paper is ``preprints'', please uncomment this parenthesis.
%\begin{landscape}

%\bibliographystyle{apj}
\bibliography{reference}

\section*{Appendix} 
\label{Appendix}
{Table \ref{table2} summarizes all the data of EUV waves and associated activities such Flares, Moreton waves, m-type II burst, SEPs, CMEs, ICMEs and $ |Dst|$ Index.}

\begin{sidewaystable}
\centering
\scriptsize
\setlength{\tabcolsep}{5pt}
%\bfseries
\caption{The parameters of the EUV waves and their associated Moreton waves, CMEs, Flares, type II radio burst, SEPs and  ICMEs. The CME onset time means the time when a CME first appears in the LASCO C2 field of view.  The ICME disturbance date and time are of the associated geomagnetic storm related to the arrival of a shock at Earth.}

\label{table2}
\begin{tabular}{ccccccccccccccc}
\toprule
 Event & Date & EUV & EUV & Flare & GOES & Moreton & m-type II & m-type II & SEP & CME &CME &ICME & ICME & $ |Dst|$ \\
 No. & YYYYMMDD & Onset & Source & Onset & Flare & Wave & Onset & Speed & Association/ & Onset &Speed & Disturbance & Speed & Index \\
 &      & Time (UT)  & location & Time (UT) & Class & (Yes/No) &  &   ({$\mathrm{km\,s^{-1}}$}) & Peak Flux (pfu)  & & ($\mathrm{km\,s^{-1}}$) & Date (UT) & ({$\mathrm{km\,s^{-1}}$}) & (nT) \\ 
\midrule
01 & 20191224 & 05:26 & N04W44  & 05:24 & No       & No & No & -- & No & 06:12 & 966 & No & -- & No \\
02 & 20200119 & 18:28 & N17E28  & 18:27 & No       & No & No & -- & No & No & -- & No & -- & No \\
03 & 20200527 & 18:05 & N32E111 & 18:05 & Backside & No & No & -- & No & 18:36 & 726 & No & -- & No \\
04 & 20200529 & 07:23 & N32E89  & 07:20 & M1.1     & No & No & -- & No & 08:00 & 337 & No & -- & No \\
05 & 20200816 & 17:18 & S31E26  & 17:12 & B1.2     & No & No & -- & No & 18:12 & 437 & No & -- & No \\
06 & 20201016 & 12:54 & S24W90  & 12:46 & C1.5     & No & No & -- & No & 13:25 & 271 & No & -- & No \\
07 & 20201016 & 15:10 & S24W90  & 14:57 & C3.5     & No & No & -- & No & 16:10 & 640 & No & -- & No \\
08 & 20201018 & 08:40 & S24W85 & 08:30  & Backside & No & No & -- & No & 09:12 & 438 & No & -- & No \\
09 & 20201101 & 19:04 & S17W82 & 19:01 & C1.6      & No & No & -- & No & 19:48 & 289 & No & -- & No \\
10 & 20201126 & 20:40 & N32E86 & 20:20 & C2.6      & No & No    & -- & No & 21:12 & 572 & No & -- & No \\
11 & 20201129 & 12:46 & S23E95 & 12:34 & M4.4      & No & 12:57 & 765 & No & 13:25 & 2077 & No & -- & No \\
12 & 20201207 & 15:48 & S25W09 & 15:46 & C7.4      & No & No & -- & No & 16:24 & 1407 & 20201210 02:10 & 440 & 14 \\
13 & 20201214 & 14:22 & S22E48 & 14:09 & C4.0      & No & 14:37 & 606 & No & 15:12 & 437 & No & -- & No \\
14 & 20201229 & 21:01 & S16W05 & 20:52 & B6.6      & No & No & -- & No & No & -- & No & -- & No \\
15 & 20210128 & 22:55 & N18W30 & 22:54 & B3.4      & No & No & -- & No & 23:36 & 300 & No & -- & No \\
16 & 20210129 & 09:48 & N18W38 & 09:41 & B2.5      & No & No & -- & No & 10:36 & 618 & No & -- & No \\
17 & 20210129 & 20:15 & N18W45 & 20:05 & B3.8      & No & No & -- & No & 20:57 & 452 & No & -- & No \\
18 & 20210226 & 07:12 & S21W46 & 07:02 & B9.5      & No & No & -- & No & 08:12 & 175 & No & -- & No \\
19 & 20210227 & 18:00 & N23W63 & 17:51 & C2.7      & No & No & -- & No & 19:00 & 118 & No & -- & No \\
20 & 20210228 & 06:38 & N20W67 & 06:36 & C3.9      & No & No & -- & No & 07:12 & 227 & No & -- & No \\
21 & 20210305 & 10:13 & N30E90 & 10:03 & B1.8      & No & No & -- & No & 11:24 & 721 & No & -- & No \\
22 & 20210305 & 16:20 & N27E90 & 16:14 & B1.7      & No & No & -- & No & 17:12 & 153 & No & -- & No \\

23 & 20210307 & 15:27 & N25E56 & 15:20 & No        & No & No & -- & No & 15:48 & 906 & No & -- & No \\
24 & 20210322 & 01:15 & N23W02 & 01:10 & B5.6      & No & No & -- & No & 02:12 & 189 & No & -- & No \\
25 & 20210327 & 20:40 & N22E05 & 20:34 & B1.2      & No & No & -- & No & No & -- & No & -- & No \\
26 & 20210417 & 16:05 & S18E90 & 15:58 & Backside  & No & 16:26 & 382 & No & 16:36 & 763 & No & -- & No \\
27 & 20210419 & 05:15 & S18E92 & 05:11 & B2.7      & No & No & -- & No & 06:00 & 230 & No & -- & No \\
28 & 20210419 & 07:49 & S18E90 & 07:34 & B1.4      & No & No & -- & No & 08:36 & 264 & No & -- & No \\
29 & 20210419 & 23:22 & S24E25 & 23:19 & M1.1      & No & 23:39 & 1202 & No & 00:24 & 427 & No & -- & No \\
30 & 20210422 & 04:20 & S25W06 & 04:10 & C3.8      & No & No & -- & No & 05:48 & 355 & No & -- & No \\
31 & 20210425 & 01:32 & S25W46 & 01:17 & C2.6      & No & No & -- & No & 02:12 & 696 & No & -- & No \\
32 & 20210503 & 20:50 & S23W90 & 20:45 & No        & No & No & -- & No & 21:36 & 309 & No & -- & No \\
33 & 20210505 & 22:28 & N17E95 & 22:25 & B1.2      & No & No & -- & No & 23:12 & 264 & No & -- & No \\
34 & 20210506 & 08:05 & N17E91 & 07:52 & B3.4      & No & No & -- & No & 08:48 & 542 & No & -- & No \\
35 & 20210507 & 18:51 & N17E78 & 18:43 & M3.9      & No & 19:02 & 816 & No & 19:24 & 754 & No & -- & No \\
36 & 20210509 & 13:43 & N15E51 & 13:38 & C4.0      & No & 14:31 & 407 & No & 15:05 & 603 & No & -- & No \\
37 & 20210517 & 15:00 & N17E83 & 14:50 & B3.4      & No & No & -- & No & 16:12 & 706 & No & -- & No \\
38 & 20210517 & 17:52 & N17E81 & 17:47 & B1.4      & No & No & -- & No & 18:24 & 899 & No & -- & No \\
39 & 20210517 & 19:06 & N17E81 & 19:00 & B4.3      & No & No & -- & No & 20:00 & 190 & No & -- & No \\
40 & 20210517 & 21:07 & N17E80 & 20:55 & B1.4      & No & No & -- & No & 22:00 & 418 & No & -- & No \\
41 & 20210517 & 22:10 & N17E80 & 22:07 & B3.5      & No & No & -- & No & 23:12 & 136 & No & -- & No \\
42 & 20210518 & 02:08 & N17E77 & 01:58 & C1.1      & No & 02:19 & 106 & No & 02:48 & 302 & No & -- & No \\
43 & 20210518 & 05:40 & N17E74 & 05:36 & B1.3      & No & No & -- & No & 06:36 & 257 & No & -- & No \\
44 & 20210518 & 08:23 & N16E73 & 08:15 & B2.3      & No & No & -- & No & 09:12 & 242 & No & -- & No \\
45 & 20210518 & 12:44 & N17E70 & 12:34 & No        & No & No & -- & No & 14:00 & 248 & No & -- & No \\

\bottomrule
\end{tabular}
\end{sidewaystable}

\begin{sidewaystable}
\centering
\scriptsize
\setlength{\tabcolsep}{5pt}
%\bfseries
\caption*{ Table \ref{table2} (continued).}
\begin{tabular}{ccccccccccccccc}
\toprule
 Event & Date & EUV & EUV & Flare & GOES & Moreton & m-type II & m-type II & SEP & CME &CME &ICME & ICME & $ |Dst|$ \\
 No. & YYYYMMDD & Onset & Source & Onset & Flare & Wave & Onset & Speed & Association/ & Onset &Speed & Disturbance & Speed & Index \\
 &      & Time (UT)  & location & Time (UT) & Class & (Yes/No) &  &   ({$\mathrm{km\,s^{-1}}$}) & Peak Flux (pfu)  & & ($\mathrm{km\,s^{-1}}$) & Date (UT) & ({$\mathrm{km\,s^{-1}}$}) & (nT) \\ 
\midrule
46 & 20210521 & 05:20 & N17E37 & 05:11   & B3.0 & No & No    & -- & No & No & -- & No & -- & No \\
47 & 20210521 & 19:27 & N18E32 & 19:15   & C4.8 & No & No    & -- & No & 20:48 & 387 & No & -- & No \\
48 & 20210522 & 02:58 & N17E27 & 02:47   & C6.1 & No & 02:54 & 1530 & No & 04:00 & 366 & No & -- & No \\
49 & 20210522 & 06:17 & N17E26 & 06:06   & C6.0 & No & 06:18 & 653 & No & 07:36 & 412 & No & -- & No \\
50 & 20210522 & 06:50 & N22E23 & 06:45   & C1.3 & No & No    & -- & No & 07:48 & 381 & No & -- & No \\
51 & 20210522 & 08:44 & N19E23 & 08:40   & No   & No & No    & -- & No & 09:48 & 292 & No & -- & No \\
52 & 20210522 & 11:19 & N22E20 & 11:17   & C2.5 & No & No    & -- & No & 12:12 & 367 & No & -- & No \\
53 & 20210522 & 15:25 & N22E17 & 15:12   & C3.2 & No & No    & -- & No & 16:24 & 355 & No & -- & No \\
54 & 20210522 & 17:11 & N20E19 & 17:03   & M1.1 & Yes & No   & -- & No & 18:00 & 605 & No & -- & No \\
55 & 20210522 & 21:34 & N22E13 & 21:30   & M1.4 & Yes & No   & -- & No & 22:12 & 459 & No & -- & No \\
56 & 20210522 & 23:15 & N20E16 & 23:05   & C6.8 & No & No    & -- & No & 00:12 & 571 & No & -- & No \\
57 & 20210523 & 09:21 & N21E07 & 09:10   & C2.4 & No & No    & -- & No & 10:12 & 392 & 20210526 12:44 & 410 & 28 \\
58 & 20210523 & 11:03 & N21E06 & 11:00   & M1.1 & Yes & No   & -- & No & 11:48 & 384 & 20210526 12:44 & 410 & 28 \\
59 & 20210523 & 16:56 & N21E02 & 16:55   & C2.2 & No & No    & -- & No & 17:24 & 518 & 20210526 12:44 & 410 & 28 \\
60 & 20210527 & 23:10 & N21W53 & 22:54   & C7.6 & No & No    & -- & No & No & -- & No & -- & No \\
61 & 20210528 & 22:33 & N21W63 & 22:19   & C9.4 & No & 23:03 & 2087 & 15 & 23:12 & 971 & No & -- & No \\
62 & 20210607 & 01:30 & S18E03 & 01:20   & B2.0 & No & No    & -- & No & No Data & -- & No & -- & No \\
63 & 20210609 & 11:54 & N27W90 & 11:54   & C1.7 & No & 12:01 & 1453 & 01 & 12:24 & 441 & No & -- & No \\
64 & 20210610 & 06:08 & S20E84 & 05:59   & C1.1 & No & No    & -- & No & 07:00 & 259 & No & -- & No \\
65 & 20210623 & 06:47 & N16E90 & 06:43   & C3.4 & No & 07:05 & 1005 & No & 07:24 & 491 & No & -- & No \\
66 & 20210625 & 13:41 & N23W82 & 13:40   & C1.7 & No & No    & -- & No & 14:12 & 433 & No & -- & No \\
67 & 20210629 & 05:18 & N18E12 & 05:14   & B6.9 & No & No    & -- & No & 06:24 & 206 & No & -- & No \\
68 & 20210703 & 07:o5 & N23W78 & 07:04   & M2.7 & No & No    & -- & No & 07:36 & 426 & No & -- & No \\
69 & 20210703 & 14:18 & N24W81 & 14:18   & X1.5 & No & 14:37 & 357 & No & 14:48 & 455 & No & -- & No \\
70 & 20210709 & 17:12 & N32W90 & 17:10   & C4.7 & No & No    & -- & No & 17:36 & 668 & No & -- & No \\
71 & 20210715 & 21:03 & Backside & 21:00 & Backside & No     & -- & No & No & 21:36 & 1476 & No & -- & No \\
72 & 20210717 & 04:51 & S15E90 & 04:48   & Backside & No     & 05:10 & 439 & No & 05:24 & 1228 & No & -- & No \\
73 & 20210720 & 16:05 & N28E67 & 16:05   & No & No & No      & -- & No & 16:36 & 873 & No & -- & No \\
74 & 20210723 & 23:58 & S15E53 & 23:54   & C4.3 & No & 00:10 & 554 & No & 00:36 & 587 & No & -- & No \\
75 & 20210725 & 04:46 & N22W25 & 04:42   & B2.5 & No & 04:52 & 711 & No & 05:48 & 299 & No & -- & No \\
76 & 20210727 & 02:02 & S13E24 & 02:00   & B1.9 & No & No    & -- & No & 03:12 & 274 & No & -- & No \\
77 & 20210826 & 18:01 & N20E14 & 17:48   & C3.0 & Yes & No   & -- & No & 18:48 & 644 & No & -- & No \\
78 & 20210826 & 23:23 & S31E23 & 23:17   & C3.9 & No & 23:29 & 691 & No & 00:24 & 197 & No & -- & No \\
79 & 20210828 & 05:54 & S28W01 & 05:39   & M4.7 & No & 06:04 & 276 & No & 06:36 & 191 & No & -- & No \\
80 & 20210829 & 08:43 & S16E20 & 08:22   & No & No & No      & -- & No & No Data & -- & No & -- & No \\
81 & 20210829 & 16:09 & S16E15 & 16:03   & No & No & No      & -- & No & 16:24 & 473 & No & -- & No \\
82 & 20210831 & 12:02 & S25W90 & 12:01   & Backside & No & No & -- & No & 12:36 & 811 & No & -- & No \\
83 & 20210904 & 21:10 & N24E06 & 21:00   & No & No & No      & -- & No & 22:24 & 227 & No & -- & No \\
84 & 20210905 & 17:53 & N28E19 & 17:50   & No & No & No      & -- & No & 19:00 & 438 & No & -- & No \\
85 & 20210907 & 23:50 & N25W30 & 23:42   & C2.3 & No & No    & -- & No & 00:24 & 379 & No & -- & No \\
86 & 20210917 & 04:05 & S30E90 & 04:05   & C5.5 & No & 04:17 & 582 & No & 04:28 & 1349 & No & -- & No \\
87 & 20210928 & 05:52 & S25W44 & 05:50   & C1.6 & No & 06:20 & 534 & No & 06:24 & 524 & 20211001 10:00 & 480 & 30 \\
88 & 20211004 & 10:04 & N32W18 & 09:53   & B2.2 & No & No    & -- & No & 11:12 & 408 & No & -- & No \\
89 & 20211009 & 06:30 & N19E08 & 06:19   & M1.6 & Yes & 06:30 & 934 & No & 07:12 & 712 & 20211012 02:26 & 440 & 65 \\
90 & 20211023 & 23:10 & S20E50 & 23:01   & B3.7 & No & No    & -- & No & 00:12 & 287 & No & -- & No \\

\bottomrule
\end{tabular}
\end{sidewaystable}

\begin{sidewaystable}
\centering
\scriptsize
\setlength{\tabcolsep}{5pt}
%\bfseries
\caption*{ Table \ref{table2} (continued).}
\begin{tabular}{ccccccccccccccc}
\toprule
  Event & Date & EUV & EUV & Flare & GOES & Moreton & m-type II & m-type II & SEP & CME &CME &ICME & ICME & $ |Dst|$ \\
 No. & YYYYMMDD & Onset & Source & Onset & Flare & Wave & Onset & Speed & Association/ & Onset &Speed & Disturbance & Speed & Index \\
 &      & Time (UT)  & location & Time (UT) & Class & (Yes/No) &  &   ({$\mathrm{km\,s^{-1}}$}) & Peak Flux (pfu)  & & ($\mathrm{km\,s^{-1}}$) & Date (UT) & ({$\mathrm{km\,s^{-1}}$}) & (nT) \\ 
\midrule
91 & 20211025 & 19:53 & N16E96 & 19:47 & C8.5 & No & No & -- & No & 20:57 & 298 & No & -- & No \\
92 & 20211026 & 02:44 & N16E94 & 02:40 & M1.3 & No & No & -- & No & No Data & -- & No & -- & No \\
93 & 20211026 & 06:01 & N16E92 & 05:56 & C7.8 & No & No & -- & No & No Data & -- & No & -- & No \\
94 & 20211026 & 09:50 & N16E89 & 09:43 & C5.0 & No & No & -- & No & 10:24 & 409 & No & -- & No \\
95 & 20211026 & 15:51 & N16E86 & 15:42 & M1.0 & No & No & -- & No & 16:36 & 612 & No & -- & No \\
96 & 20211026 & 23:15 & N18E80 & 23:10 & C2.1 & No & No & -- & No & 00:24 & 329 & No & -- & No \\
97 & 20211028 & 15:17 & S26E02 & 15:17 & X1.0 & Yes & 15:29 & 1457 & 29 & 15:48 & 1519 & No & -- & No \\
98 & 20211030 & 19:46 & N28W13 & 19:49 & B6.6 & No & No & -- & No & No Data & -- & No & -- & No \\
99 & 20211101 & 01:08 & S27W45 & 00:57 & M1.5 & No & 01:29 & 626 & No & 02:00 & 753 & 20211103 19:42 & 650 & 105 \\
100 & 20211101 & 20:56 & S28W56 & 20:54 & C4.0 & No & No & -- & No & 21:24 & 830 & 20211103 19:42 & 650 & 105 \\
101 & 20211102 & 02:00 & N20E05 & 01:50 & M1.7 & No & No & -- & No & 02:48 & 1473 & 20211103 19:42& 650 & 105 \\
102 & 20211103 & 21:01 & S26W89 & 20:56 & C5.2 & No & No & -- & No & 21:36 & 510 & No & -- & No \\
103 & 20211109 & 16:11 & N16W90 & 16:07 & M2.0 & No & No & -- & 02 & 16:48 & 1350 & No & -- & No \\
104 & 20211204 & 23:59 & S26W90 & 23:44 & C7.1 & No & No & -- & No & 00:24 & 507 & No & -- & No \\
105 & 20211205 & 07:06 & S26W90 & 06:58 & M1.4 & No & No & -- & No & 07:36 & 628 & No & -- & No \\
106 & 20211205 & 18:32 & S26W90 & 18:32 & C1.6 & No & No & -- & No & 19:00 & 563 & No & -- & No \\
107 & 20211206 & 05:27 & S26W90 & 05:26 & B9.0 & No & No & -- & No & 05:48 & 992 & No & -- & No \\
108 & 20211206 & 06:18 & N17W76 & 06:16 & C1.3 & No & No & -- & No & 07:48 & 334 & No & -- & No \\
109 & 20211218 & 17:29 & S19E80 & 17:27 & C6.3 & No & No & -- & No & 18:00 & 588 & No & -- & No \\
110 & 20211220 & 03:18 & S19E11 & 03:08 & C4.4 & No & No & -- & No & 04:12 & 294 & No & -- & No \\
111 & 20211220 & 11:22 & S20W01 & 11:12 & M1.8 & No & 11:22 & 350 & No & 12:24 & 359 & No & -- & No \\
112 & 20211221 & 19:04 & S21W06 & 19:02 & C4.9 & Yes & No & -- & No & 19:48 & 232 & No & -- & No \\
113 & 20211225 & 09:03 & S22W55 & 09:00 & C1.8 & No & No & --& No & 10:12 & 192 & No & -- & No \\
114 & 20211228 & 04:05 & N20W20 & 03:50 & M1.8 & No & No & -- & No & 05:00 & 306 & No & -- & No \\
115 & 20220109 & 22:40 & N33E90 & 22:37 & Backside & No & 22:52 & 575 & No & 23:12 & 1283 & No & -- & No \\
116 & 20220112 & 04:20 & N33E120 & 04:10 & Backside & No & 04:26 & 535 & No & 04:36 & 1586 & No & -- & No \\
117 & 20220117 & 13:57 & N24W40 & 13:48 & C2.8 & No & No & -- & No & 14:48 & 511 & No & -- & No \\
118 & 20220118 & 17:19 & N09W55 & 17:16 & M1.5 & No & 17:31 & 143 & No & 17:48 & 1014 & No & -- & No \\
119 & 20220120 & 05:41 & N09W76 & 05:41 & M5.5 & No & 05:57 & 329 & 22 & 06:12 & 1431 & No & -- & No \\
120 & 20220129 & 22:50 & N18E07 & 22:45 & M1.1 & No & No & -- & No & 23:36 & 530 & 20220201 22:20 & 460  & 66 \\
121 & 20220206 & 13:04 & S20W07 & 12:52 & C3.1 & No & No & -- & No & 14:00 & 334 & 20220209 21:00& 400 & 60 \\
122 & 20220211 & 12:06 & S24E26 & 11:53 & No & No & No & --& No & 13:48 & 165 & No & -- & No \\
123 & 20220212 & 08:30 & S18W80 & 08:25 & M1.4 & No & 08:33 & 642 & No & 09:12 & 280 & No & -- & No \\
124 & 20220212 & 17:12 & S18W85 & 17:12 & Backside & No & No & -- & No & 17:36 & 866 & No & -- & No \\
125 & 20220215 & 21:46 & N40E85 & 21:44 & Backside & No & No & -- & No & 22:12 & 1905 & No & -- & No \\
126 & 20220217 & 20:45 & N24E38 & 20:40 & C1.2 & No & No & -- & No & 21:48 & 423 & No & -- & No \\
127 & 20220218 & 16:08 & N25E28 & 16:05 & B8.4 & No & No & -- & No & 17:48 & 589 & No & -- & No \\
128 & 20220219 & 12:11 & N25E16 & 11:56 & B3.3 & No & No & -- & No & 13:36 & 623 & No & -- & No \\
129 & 20220302 & 17:36 & N18E30 & 17:31 & M2.0 & No & 17:36 & 690 & No & 18:12 & 266 & No & -- & No \\
130 & 20220303 & 21:51 & N18E14 & 21:44 & C3.3 & No & No & -- & No & No Data & -- & No & -- & No \\
131 & 20220306 & 22:40 & N18W35 & 22:30 & No & No & No & -- & No & 00:12  & 241 & No & -- & No \\
132 & 20220312 & 00:05  & S20E60 & 23:55 & C1.4 & No & No & -- & No & 00:48  & 879 & No & -- & No \\
133 & 20220314 & 17:10 & S10W85 & 17:08 & Backside & No & 17:20 & 632 & 01 & 17:36 & 740 & No & -- & No \\
134 & 20220316 & 12:42 & N24E19 & 12:37 & No & No & No & -- & No & 13:25 & 491 & No & -- & No \\
135 & 20220320 & 07:43 & N19E14 & 07:30 & C4.6 & No & 07:43 & 188 & No & No & -- & No & -- & No \\
\bottomrule
\end{tabular}
\end{sidewaystable}

\begin{sidewaystable}
\centering
\scriptsize
\setlength{\tabcolsep}{5pt}
%\bfseries
\caption*{ Table \ref{table2} (continued).}
\begin{tabular}{ccccccccccccccc}
\toprule
 Event & Date & EUV & EUV & Flare & GOES & Moreton & m-type II & m-type II & SEP & CME &CME &ICME & ICME & $ |Dst|$ \\
 No. & YYYYMMDD & Onset & Source & Onset & Flare & Wave & Onset & Speed & Association/ & Onset &Speed & Disturbance & Speed & Index \\
 &      & Time (UT)  & location & Time (UT) & Class & (Yes/No) &  &   ({$\mathrm{km\,s^{-1}}$}) & Peak Flux (pfu)  & & ($\mathrm{km\,s^{-1}}$) & Date (UT) & ({$\mathrm{km\,s^{-1}}$}) & (nT) \\ 
\midrule
136 & 20220321 & 05:31 & Backside & 05:30 & Backside & No & 05:49 & 394 & 06 & 06:00 & 916 & No &  -- & No  \\
137 & 20220325 & 05:11 & S18E39 & 05:02 & M1.4 & No & 05:14 & 959 & No & 05:48 & 495 & No & -- & No \\
138 & 20220328 & 11:21 & N15W04 & 11:15 & M4.0 & No & 11:23 & 1259 & 19 & 12:00 & 702 & 20220331 02:24 & 490 & 19 \\
139 & 20220328 & 19:12 & N16W06 & 19:00 & M1.0 & No & 20:04 & 1605 & No & 20:24 & 905 & 20220331 02:24 & 490 & 19 \\
140 & 20220330 & 17:30 & N15W32 & 17:21 & X1.3 & Yes & 17:32 & 1424 & 10 & 18:00 & 641 & No & -- & No  \\
141 & 20220331 & 17:28 & S14E25 & 17:27 & C2.1 & No & No & -- & No & No Data & -- & No & -- & No \\
142 & 20220331 & 18:27 & N14W46 & 18:17 & M9.6 & Yes & 18:29 & 772 & No & 19:12 & 489 & No & -- & No \\
143 & 20220331 & 20:48 & S15E24 & 20:48 & C2.6 & No & No & -- & No & 21:24 & 449 & No & -- & No \\
144 & 20220402 & 13:02 & N16W66 & 12:56 & M3.9 & No & 13:23& 956 & 32 & 13:36 & 1433 & No & -- & No \\
145 & 20220411 & 05:00 & S14E06 & 04:59 & C1.6 & No & No & -- & No & 05:48 & 940 & 20220414 09:00 & 520 & 81 \\
146 & 20220417 & 03:12 & N12E84 & 03:10 & X1.1 & No & 03:28 & 724 & No & 03:48 & 895 & No & -- & No \\
147 & 20220420 & 01:27 & S33W92 & 01:20 & M7.2 & No & No & -- & No & 02:00 & 542 & No & -- & No \\
148 & 20220420 & 03:44 & S33W93 & 03:41 & X2.2 & No & 03:55 & 1630 & No & 04:12 & 1001 & No & -- & No \\
149 & 20220421 & 01:50 & N22E23 & 01:47 & M9.6 & No & 01:57 & 1132 & No & 02:24 & 357 & No & -- & No \\
150 & 20220510 & 13:54 & S29W04 & 13:50 & X1.5 & Yes & 13:58 & 718 & No & 14:36 & 415 & No & -- & No \\
151 & 20220525 & 18:08 & S19W41 & 18:06 & M1.3 & Yes & 18:17 & 1638 & No & 18:36 & 1134 & No & -- & No \\
152 & 20220602 & 06:00 & S17E93 & 05:58 & C1.2 & No & No & -- & No & 06:48 & 581 & No & -- & No \\
153 & 20220626 & 08:10 & S12E09 & 08:00 & C2.4 & No & No & -- & No & No & -- & No & -- & No \\
154 & 20220704 & 13:32 & N17E36 & 13:23 & C5.1 & No & 13:32 & 1828 & No & 14:24 & 564 & No & -- & No \\
155 & 20220705 & 04:02 & S20W18 & 03:59 & C9.8 & No & 04:16 & 514 & No & 04:36 & 747 & 20220708 02:00 & 400 & 37 \\
156 & 20220708 & 20:08 & N20E40 & 20:04 & M2.5 & No & No & -- & No & 20:36 & 1223 & No & -- & No \\
157 & 20220709 & 13:27 & S22W88 & 13:24 & C8.5 & No & 13:50 & 828 & 04 & 13:48 & 1034 & No & -- & No \\
158 & 20220714 & 21:46 & N15E69 & 21:42 & M2.8 & Yes & No & -- & No & 22:12 & 517 & No & -- & No \\
159 & 20220721 & 01:05 & N13W02 & 01:02 & C5.6 & No & 01:21 & 1063 & No & 01:48 & 875 & No & -- & No \\
160 & 20220803 & 16:59 & S13W22 & 16:54 & C6.9 & No & 17:05 & 163 & No & 18:00 & 400 & No & -- & No \\
161 & 20220815 & 14:37 & S22W00 & 14:27 & M1.0 & No & No & -- & No & 15:24 & 467 & No & -- & No \\
162 & 20220815 & 16:54 & S22W00 & 16:40 & M2.7 & Yes & No & --& No & 17:48 & 599 & No & -- & No \\
163 & 20220816 & 07:55 & S22W11 & 07:52 & M5.0 & Yes & No & -- & No & 08:48 & 609 & No & -- & No \\
164 & 20220819 & 04:20 & S27W48 & 04:14 & M1.6 & No & 04:35 & 694 & No & 04:49 & 832 & No & -- & No \\
165 & 20220819 & 20:25 & S26W55 & 20:19 & C6.7 & No & 20:33 & 318 & No & 21:12 & 332 & No & -- & No \\
166 & 20220827 & 01:57 & S30W62 & 01:52 & M4.8 & No & 02:12 & 657 & No & 02:24 & 1284 & No & -- & No \\
167 & 20220828 & 15:54 & S30W85 & 15:48 & M6.7 & No & No & -- & No & 16:12 & 1232 & No & -- & No \\
168 & 20220831 & 04:13 & S21W05 & 04:04 & C3.3 & No & No & -- & No & 04:48 & 816 & No & -- & No \\
169 & 20220923 & 17:49 & N16E84 & 17:48 & M1.7 & No & 17:56 & 620 & No & 18:12 & 1776 & No & -- & No \\
170 & 20221002 & 02:09 & N17W39 & 02:07 & M8.7 & No & No & -- & No & 02:36 & 1222 & No & -- & No \\
171 & 20221002 & 20:10 & N17W49 & 20:10 & X1.0 & Yes & 20:24 & 1157 & 01 & 20:36 & 1086 & No & -- & No \\
172 & 20221004 & 12:56 & N17W75 & 12:48 & M1.6 & No & No & -- & No & 13:25 & 1089 & No & -- & No \\
173 & 20221011 & 08:45 & N24W35 & 08:36 & M3.9 & No & 08:47 & 967 & No & 09:12 & 252 & No & -- & No \\
174 & 20221106 & 00:05 & N12E56 & 23:59 & M5.2 & No & 00:11 & 927 & No & 00:48 & 422 & No & -- & No \\
175 & 20221109 & 19:52 & S27E62 & 19:46 & C4.3 & No & 20:00 & 1242 & No & 20:24 & 228 & No & -- & No \\
176 & 20221118 & 09:55 & N22W35 & 09:50 & No & No & No & -- & No & 10:24 & 225 & No & -- & No \\
177 & 20221118 & 11:26 & N20W35 & 11:23 & C5.1 & No & No & -- & No & No Data & -- & No & -- & No \\
178 & 20221118 & 18:47 & N20W39 & 18:44 & C4.6 & No & No & -- & No & 20:00 & 517 & No & -- & No \\
179 & 20221119 & 06:05 & N20W46 & 06:05 & C8.3 & No & No & -- & No & 07:24 & 151 & No & -- & No \\
180 & 20221119 & 12:48 & N20W50 & 12:42 & M1.6 & No & 13:01 & 930 & No & 13:25 & 422 & No & -- & No \\

\bottomrule
\end{tabular}
\end{sidewaystable}

\begin{sidewaystable}
\centering
\scriptsize
\setlength{\tabcolsep}{5pt}
%\bfseries
\caption*{ Table \ref{table2} (continued).}
\begin{tabular}{ccccccccccccccc}
\toprule
 Event & Date & EUV & EUV & Flare & GOES & Moreton & m-type II & m-type II & SEP & CME &CME &ICME & ICME & $ |Dst|$ \\
 No. & YYYYMMDD & Onset & Source & Onset & Flare & Wave & Onset & Speed & Association/ & Onset &Speed & Disturbance & Speed & Index \\
 &      & Time (UT)  & location & Time (UT) & Class & (Yes/No) &  &   ({$\mathrm{km\,s^{-1}}$}) & Peak Flux (pfu)  & & ($\mathrm{km\,s^{-1}}$) & Date (UT) & ({$\mathrm{km\,s^{-1}}$}) & (nT) \\ 
\midrule
181 & 20221215 & 07:56 & S20W48 & 07:54 & M1.0 & No & No & -- & No & 09:12 & 589 & No & -- & No \\
182 & 20221230 & 15:28 & N18E08 & 15:24 & M1.4 & Yes & No & -- & No & No Data & -- & 20230103 21:00 & 380 & 61 \\
183 & 20230103 & 06:14 & S18E104 & 06:14 & C3.2 & No & No & -- & No & 06:36 & 1129 & No & -- & No \\
184 & 20230217 & 19:44 & N25E67 & 19:38 & X2.2 & No & 19:57 & 2407 & No & 20:12 & 1315 & No & -- & No \\
185 & 20230225 & 18:50 & N24W45 & 18:40 & M6.3 & No & 19:23 & 3133 & 58 & 19:24 & 1170 & 20230227 10:00 & 650 & 132 \\
186 & 20230302 & 21:05 & N21W64 & 21:05 & M3.8 & Yes & No & -- & No & 21:36 & 244 & No & -- & No \\
187 & 20230303 & 17:43 & N21W76 & 17:41 & X2.1 & No & 17:53 & 247 & No & 18:12 & 709 & No & -- & No \\
188 & 20230304 & 06:57 & N16W38 & 06:54 & M1.0 & Yes & No & -- & No & 07:24 & 1178 & No & -- & No \\
189 & 20230304 & 15:05 & N19W87 & 15:03 & M5.2 & No & 15:31 & 815 & No & 15:36 & 933 & No & -- & No \\
190 & 20230306 & 02:21 & N18W63 & 02:08 & M5.8 & No & No & -- & No & 03:12 & 873 & No & -- & No \\
191 & 20230317 & 06:47 & S19E80 & 06:45 & C6.4 & No & No & -- & No & 07:14 & 452 & No & -- & No \\
192 & 20230318 & 00:35 & S19E72 & 00:27 & C3.8 & No & No & -- & No & 01:10 & 557 & No & -- & No \\
193 & 20230318 & 07:06 & S19E70 & 07:04 & C9.4 & No & No & -- & No & 07:29 & 763 & No & -- & No \\
194 & 20230321 & 10:20 & N20E11 & 10:16 & C1.7 & No & No & -- & No & 11:24 & 324 & No & -- & No \\
195 & 20230330 & 07:22 & S22W77 & 07:21 & M5.4 & No & 07:36 & 279 & 06 & 07:48 & 487 & No & -- & No \\
196 & 20230401 & 02:52 & S22W100 & 02:49 & C6.7 & No & No & -- & No & 03:12 & 1006 & No & -- & No \\
197 & 20230415 & 22:49 & N25W36 & 22:35 & No & No & No & -- & No & 00:12 & 390 & No & -- & No \\
198 & 20230421 & 17:48 & S23W10 & 17:44 & M1.7 & No & 17:56 & 580 & 26 & 18:12 & 1284 & 20230423 17:38 & 520 & 213 \\
199 & 20230505 & 07:32 & N14E32 & 07:30 & M1.2 & No & 07:54 & 997 & No & 08:00 & 770 & No & -- & No \\
200 & 20230506 & 21:50 & S08E42 & 21:42 & C4.7 & No & No & -- & No & 22:36 & 174 & No & -- & No \\
201 & 20230507 & 21:55 & N16W06 & 21:40 & M1.5 & No & No & -- & 38 & 23:12 & 1075 & 20230509 22:50 & 550 & 19 \\
202 & 20230509 & 03:53 & N13W26 & 03:42 & M6.5 & No & 03:52 & 774 & No & 04:24 & 416 & No & -- & No \\
203 & 20230509 & 18:30 & N15W31 & 18:20 & M4.2 & No & 18:40 & 1892 & 83 & 19:00 & 1209 & 20230512 06:33 & 480 & 16 \\
204 & 20230511 & 08:54 & S06W41 & 08:47 & M2.2 & Yes & 08:57 & 419 & No & 09:24 & 365 & No & -- & No \\
205 & 20230620 & 16:51 & S17E73 & 16:42 & X1.1 & No & 17:01 & 1027 & No & 17:12 & 1113 & No & -- & No \\
206 & 20230622 & 23:38 & S21E41 & 23:29 & M4.9 & No & 23:45 & 534 & No & No Data & -- & No & -- & No \\
207 & 20230623 & 03:33 & N19W27 & 03:20 & C4.6 & No & No & -- & No & 04:14 & 302 & No & -- & No \\
208 & 20230625 & 21:35 & N10W30 & 21:30 & No & No & No & -- & No & 22:12 & 1126 & No & -- & No \\
209 & 20230703 & 21:03 & N21E34 & 21:02 & C8.9 & No & 21:09 & 716 & No & 21:48 & 208 & No & -- & No \\
210 & 20230710 & 11:15 & S05W90 & 11:05 & Backside & No & 11:46 & 580 & No & 12:00 & 945 & No & -- & No \\
211 & 20230716 & 04:24 & N25W90 & 04:20 & Backside & No & No & -- & 18 & 04:42 & 1970 & No & -- & No \\
212 & 20230716 & 13:18 & N28W90 & 13:15 & Backside & No & No & -- & No & 13:41 & 1829 & No & -- & No \\
213 & 20230716 & 17:37 & S24W59 & 17:36 & M4.0 & No & No & -- & No & 17:53 & 1239 & No & -- & No \\
214 & 20230717 & 23:16 & S21W69 & 23:14 & M5.0 & No & No & -- & No & 23:36 & 1385 & No & -- & No \\
215 & 20230719 & 17:06 & S18W88 & 17:04 & M3.8 & No & 17:16 & 429 & No & 17:36 & 716 & No & -- & No \\
216 & 20230725 & 21:12 & S10E40 & 21:08 & M1.6 & Yes & No & -- & No & 22:12 & 211 & No & -- & No \\
217 & 20230728 & 15:43 & N26W87 & 15:39 & M4.1 & No & 15:52 & 550 & 154 & 16:00 & 1896 & No & -- & No \\
218 & 20230805 & 21:46 & N11W77 & 21:45 & X1.6 & No & 22:15 & 772 & No & 22:12 & 1647 & No & -- & No \\
219 & 20230806 & 18:23 & N08W89 & 18:20 & M5.5 & No & No & -- & No & 18:48 & 1182 & No & -- & No \\
220 & 20230807 & 20:26 & N10W102 & 20:27 & X1.5 & No & 20:38 & 368 & 47 & 20:48 & 1851 & No & -- & No \\
221 & 20230817 & 02:13 & S23W86 & 02:13 & C3.6 & No & 02:29 & 253 & No & 02:48 & 623 & No & -- & No \\
222 & 20230901 & 02:55 & N13W73 & 02:40 & M1.2 & No & 03:08 & 743 & 25 & 03:24 & 1339 & No & -- & No \\
223 & 20230905 & 22:20 & S21W08 & 22:13 & C3.9 & No & No & -- & No & No & -- & No & -- & No \\
224 & 20230919 & 20:01 & N07E45 & 20:01 & M4.0 & Yes & No & -- & No & 20:36 & 483 & No & -- & No \\
225 & 20230920 & 14:18 & N07E35 & 14:11 & M8.2 & Yes & 14:25 & 1054 & No & 15:05 & 199 & No & -- & No \\

\bottomrule
\end{tabular}
\end{sidewaystable}

\begin{sidewaystable}
\centering
\scriptsize
\setlength{\tabcolsep}{5pt}
%\bfseries
\caption*{ Table \ref{table2} (continued).}
\begin{tabular}{ccccccccccccccc}
\toprule
 Event & Date & EUV & EUV & Flare & GOES & Moreton & m-type II & m-type II & SEP & CME &CME &ICME & ICME & $ |Dst|$ \\
 No. & YYYYMMDD & Onset & Source & Onset & Flare & Wave & Onset & Speed & Association/ & Onset &Speed & Disturbance & Speed & Index \\
 &      & Time (UT)  & location & Time (UT) & Class & (Yes/No) &  &   ({$\mathrm{km\,s^{-1}}$}) & Peak Flux (pfu)  & & ($\mathrm{km\,s^{-1}}$) & Date (UT) & ({$\mathrm{km\,s^{-1}}$}) & (nT) \\ 
\midrule
226 & 20230921 & 12:53 & N07E22 & 12:42 & M8.7 & Yes & No & -- & No & 13:36 & 593 & No & -- & No \\
227 & 20230922 & 19:27 & S21W63 & 19:20 & C9.8 & No & 19:28 & 483 & No & 20:12 & 221 & No & -- & No \\
228 & 20231002 & 12:40 & N18E69 & 12:35 & M1.9 & Yes & No & -- & No & 13:24 & 691 & No & -- & No \\
229 & 20231007 & 18:07 & S11E40 & 17:57 & M1.7 & No & 18:07 & 598 & No & 19:12 & 233 & No & -- & No \\
230 & 20231008 & 03:38 & S09E32 & 03:34 & C6.6 & Yes & No & --  & No & 04:12 & 446 & No & -- & No \\
231 & 20231008 & 14:05 & S09E26 & 14:01 & C2.4 & No & No & -- & No & 15:37 & 582 & No & -- & No \\
232 & 20231014 & 05:08 & S10W44 & 05:03 & C9.8 & No & No & -- & No & 06:12 & 211 & No & -- & No \\
233 & 20231017 & 04:37 & S19W32 & 04:25 & C1.1 & No & No & -- & No & 05:24 & 610 & No & -- & No \\
234 & 20231026 & 22:55 & N25E90 & 22:47 & M1.4 & No & 23:18 & 961 & No & 23:14 & 1238 & No & -- & No \\
235 & 20231123 & 02:19 & N28E85 & 02:06 & Backside & No & No & -- & No & 03:12 & 982 & No & -- & No \\
236 & 20231123 & 05:25 & N28E85 & 05:15 & Backside & No & No & -- & No & 06:00 & 990 & No & -- & No \\
237 & 20231127 & 18:22 & N18E10 & 18:15 & C6.7 & No & No & -- & No & 20:00 & 307 & No & -- & No \\
238 & 20231128 & 19:23 & S15W01 & 19:35 & M9.8 & Yes & 19:34 & 694 & No & 20:12 & 741 & 20231201 09:26 & 460 & 107 \\
239 & 20231214 & 16:44 & N04W53 & 16:44 & X2.8 & Yes & 17:00 & 2118 & 13 & 17:12 & 948 & 20231217 07:39 & 470 & 77\\
240 & 20231231 & 21:42 & N04E79 & 21:36 & X5.0 & No & No & -- & 01 & 22:00 & 2852 & No & -- & No \\
241 & 20240120 & 08:35 & S16E17 & 08:35 & C3.5 & No & No & -- & No & 09:12 & 850 & No & -- & No \\
242 & 20240129 & 03:55 & N28W86 & 03:54 & M6.8 & No & 04:07 & 512 & 137 & 04:24 & 1554 & No & -- & No \\
243 & 20240207 & 03:07 & S34W70 & 03:04 & M5.1 & Yes & No & -- & No & 03:24 & 1452 & No & -- & No \\
244 & 20240208 & 23:14 & S35W98 & 23:13 & M3.9 & No & No & -- & No & 23:48 & 936 & No & -- & No \\
245 & 20240209 & 12:54 & S35W105 & 12:53 & X3.3 & No & No & -- & 187 & 13:25 & 2782 & No & -- & No \\
246 & 20240210 & 03:17 & S20E90 & 03:14 & M3.4 & No & No & -- & No & 03:48 & 1418 & No & -- & No \\
247 & 20240210 & 22:57 & S12W12 & 22:56 & M9.0 & Yes & 23:05 & 1856 & No & 23:36 & 817 & No & -- & No \\
248 & 20240212 & 06:17 & S35W90 & 06:17 & Backside & No & No & -- & No & 06:36 & 2792 & No & -- & No \\
249 & 20240214 & 03:48 & Backside & 03:45 & Backside & No & 03:55 & 1081 & No & 04:12 & 2191 & No & -- & No \\
250 & 20240216 & 06:38 & S19W82 & 06:38 & X2.5 & No & 06:53 & 753 & 07 & 07:12 & 617 & No & -- & No \\
251 & 20240221 & 01:41 & N40E90 & 01:40 & Backside & No & 02:01 & 627 & No & 02:12 & 1014 & No & -- & No \\
252 & 20240222 & 22:10 & N17E27 & 22:08 & X6.3 & No & No & -- & No & 23:00 & 204 & No & -- & No \\
253 & 20240310 & 12:05 & S13W38 & 12:00 & M7.4 & No & 12:08 & 714 & No & 12:48 & 307 & No & -- & No \\
254 & 20240323 & 00:57 & N17W03 & 00:55 & X1.1 & No & 01:09 & 791 & 956 & 01:25 & 1470 & 20240324 14:34 & 700 & 128 \\
255 & 20240503 & 02:08 & N25E06 & 02:05 & X1.6 & No & 02:20 & 944 & No & 02:48 & 808 & No & -- & No \\
256 & 20240503 & 20:20 & N09E33 & 20:12 & C3.9 & No & 20:22 & 392 & No & 21:17 & 371 & No & -- & No \\
257 & 20240504 & 06:17 & N26W09 & 06:02 & M9.1 & No & 06:25 & 950 & No & 06:36 & 576 & No & -- & No \\
258 & 20240506 & 06:03 & N26W36 & 05:48 & X4.5 & No & No & -- & No & 07:12 & 327 & No & -- & No \\
259 & 20240508 & 04:42 & S22W11 & 04:37 & X1.0 & No & 05:01 & 887 & No & 05:36 & 530 & 20240510 17:05 & 800 & 406 \\
260 & 20240508 & 11:32 & S20W11 & 11:26 & M8.6 & No & No & -- & No & 12:24 & 677 & 20240510 17:05 & 800 & 406 \\
261 & 20240509 & 08:57 & S20W24 & 08:45 & X2.2 & Yes & No & -- & No & 09:24 & 1280 & 20240510 17:05 & 800 & 406 \\
262 & 20240509 & 17:32 & S17W28 & 17:23 & X1.1 & No & 17:32 & 303 & No & 18:23 & 663 & 20240510 17:05 & 800 & 406 \\
263 & 20240510 & 06:42 & S17W34 & 06:27 & X3.9 & Yes & 06:50 & 489 & 208 & 07:12 & 953 & 20240512 09:17 & 660 & -- \\
264 & 20240511 & 01:10 & S17W44 & 01:10 & X5.8 & No & 01:13 & 564 & 116 & 01:36 & 1614 & 20240512 09:17 & 660 & -- \\
265 & 20240514 & 16:40 & S17W95 & 16:40 & X8.7 & No & 16:50 & 279 & No & 17:00 & 2010 & No & -- & No \\
266 & 20240514 & 17:20 & N17E72 & 17:20 & M4.4 & No & 17:30 & 610 & No & 17:48 & 1407 & No & -- & No \\
267 & 20240515 & 08:15 & S17W103 & 08:13 & X3.4 & No & No & -- & No & 08:36 & 1648 & No & -- & No \\
268 & 20240515 & 14:22 & S13E83 & 14:20 & X2.9 & No & 14:28 & 559 & No & 14:53 & 1209 & No & -- & No \\
269 & 20240515 & 20:31 & S17W110 & 20:30 & C5.2 & No & No & -- & No & 20:56 & 1250 & No & -- & No \\
270 & 20240520 & 05:15 & Backside & 05:15 & Backside & Yes & No & -- & No & 05:36 & 1459 & No & -- & No \\

\bottomrule
\end{tabular}
\end{sidewaystable}

\begin{sidewaystable}
\centering
\scriptsize
\setlength{\tabcolsep}{5pt}
%\bfseries
\caption*{ Table \ref{table2} (continued).}
\begin{tabular}{ccccccccccccccc}
\toprule
 Event & Date & EUV & EUV & Flare & GOES & Moreton & m-type II & m-type II & SEP & CME &CME &ICME & ICME & $ |Dst|$ \\
 No. & YYYYMMDD & Onset & Source & Onset & Flare & Wave & Onset & Speed & Association/ & Onset &Speed & Disturbance & Speed & Index \\
 &      & Time (UT)  & location & Time (UT) & Class & (Yes/No) &  &   ({$\mathrm{km\,s^{-1}}$}) & Peak Flux (pfu)  & & ($\mathrm{km\,s^{-1}}$) & Date (UT) & ({$\mathrm{km\,s^{-1}}$}) & (nT) \\ 
\midrule
271 & 20240527 & 06:54 & S18E89 & 06:49 & X2.8 & No & 06:59 & 1135 & No & 07:24 & 1291 & No & -- & No \\
272 & 20240601 & 18:31 & S20E25 & 18:30 & M7.3 & No & 19:20 & 467 & No & 19:00 & 946 & No & -- & No \\
273 & 20240608 & 01:11 & S18W69 & 01:10 & M9.7 & Yes & 01:28 & 894 & 1036 & 01:48 & 1427 & No & -- & No \\
274 & 20240610 & 10:10 & S17W101 & 09:50 & X1.5 & No & No & -- & No & 10:48 & 972 & No & -- & No \\
275 & 20240610 & 18:06 & S17W106 & 18:02 & M9.5 & No & 18:29 & 754 & No & 18:36 & 1266 & No & -- & No \\
276 & 20240611 & 22:30 & S17W112 & 22:23 & Backside & No & 22:51 & 528 & 09 & 23:06 & 1988 & No & -- & No \\
277 & 20240621 & 23:21 & S14E90 & 23:20 & C6.6 & No & 23:40 & 590 & No & 23:48 & 1388 & No & -- & No \\
278 & 20240703 & 07:32 & S18W61 & 07:29 & M1.5 & No & 07:42 & 1254 & No & 08:00 & 467 & No & -- & No \\
279 & 20240727 & 02:32 & S07E25 & 02:21 & M4.2 & No & No & No & No & 03:48 & 1413 & No & -- & No \\
280 & 20240728 & 01:49 & S07E11 & 01:53 & M9.9 & No & No & -- & No & 02:24 & 522 & 20240731 11:28 & 430 & 25 \\
281 & 20240729 & 12:51 & S11W50 & 12:47 & M8.7 & Yes & 13:00 & 567 & No & 13:25 & 956 & 20240731 11:28 & 430 & 25 \\
282 & 20240731 & 17:58 & S15W72 & 17:57 & M5.4 & No & 18:19 & 682 & No & 18:36 & 1118 & No & -- & No \\
283 & 20240805 & 13:20 & S10W88 & 13:20 & X1.7 & No & 13:31 & 171 & 02 & 13:48 & 1155 & No & -- & No \\
284 & 20240808 & 19:06 & S04W22 & 19:01 & X1.3 & No & 19:29 & 1026 & No & 19:48 & 789 & 20240810 12:50 & 460 & 188 \\
285 & 20240814 & 06:18 & N13E04 & 06:04 & X1.1 & Yes & 06:49 & 516 & No & 07:36 & 570 & 20240817 14:23 & 450 & 20 \\
286 & 20240901 & 11:37 & S23E86 & 11:36 & M5.5 & No & No & -- & No & 12:00 & 1381 & No &-- & No \\
287 & 20240909 & 04:57 & S10E90 & 04:50 & Backside & No & 05:10 & 668 & No & 05:24 & 1522 & No & -- & No \\
288 & 20240914 & 15:15 & S15E56 & 15:13 & X4.5 & No & No & -- & 06 & 15:36 & 2366 & No & -- & No \\
289 & 20241001 & 22:00 & S16E16 & 21:58 & X7.1 & No & 22:17 & 1246 & No & 22:36 & 598 & No & -- & No \\
290 & 20241003 & 12:10 & S15W03 & 12:08 & X9.0 & No & 12:18 & 573 & No & 12:48 & 821 & 20241006 07:39 & 480 & 96 \\
291 & 20241003 & 20:02 & S07W45 & 20:00 & M6.7 & Yes & 20:12 & 1241 & No & 20:36 & 1013 & 20241008 00:00 & 420 & 148 \\
292 & 20241009 & 01:40 & N13W08 & 01:25 & X1.8 & Yes & No & -- & 1810 & 02:12 & 1435 & 20241010 15:14 & 690 & 333 \\
293 & 20241024 & 03:27 & S16E76 & 03:27 & X3.3 & Yes & 03:46 & 610 & No & 03:48 & 2385 & No & -- & No \\
294 & 20241026 & 06:13 & S16E72 & 06:12 & X1.8 & No & 06:29 & 486 & 364 & 06:36 & 1914 & No & -- & No \\
295 & 20241104 & 01:04 & S06E41 & 00:52 & M3.8 & No & 01:28 & 307 & No & 01:36 & 960 & No & -- & No \\
296 & 20241121 & 17:50 & S09W120 & 17:48 & Backside & No & No & -- & 125 & 18:12 & 1436 & No & -- & No \\
297 & 20241125 & 20:35 & S11W25 & 20:24 & M1.9 & No & 20:40 & 591 & No & 21:24 & 574 & 20241129 03:05 & 390 & 09 \\
298 & 20241205 & 10:33 & S21W90 & 10:32 & M2.5 & No & 10:40 & 456 & No & 11:00 & 381 & No & -- & No \\
299 & 20241208 & 08:55 & S09W53 & 08:50 & X2.2 & No & 09:06 & 360 & No & 09:24 & 728 & No & -- & No \\
300 & 20241215 & 14:03 & S35E90 & 14:02 & Backside & No & 14:27 & 320 & No & 14:36 & 899 & No & -- & No \\
301 & 20241221 & 19:20 & N15E90 & 19:18 & Backside & No & 19:39 & 1314 & No & 19:48 & 607 & No & -- & No \\
302 & 20241223 & 10:58 & S19E32 & 10:56 & M8.9 & Yes & 11:11 & 731 & No & 11:24 & 688 & No & -- & No \\
303 & 20241224 & 08:43 & S19E18 & 08:28 & M4.1 & No & 08:41 & 455 & No & 09:24 & 255 & No & -- & No \\
304 & 20250106 & 16:22 & N11E25 & 16:12 & M4.8 & No & 16:21 & 445 & No & 17:12 & 314 & No & -- & No \\
305 & 20250107 & 22:37 & S21W101 & 22:35 & M1.1 & No & No & -- & No & 23:00 & 446 & No & -- & No \\
306 & 20250121 & 10:15 & S17E24 & 10:08 & M3.3 & No & 10:27 & 565 & No & 10:48 & 482 & No & -- & No \\
307 & 20250122 & 10:45 & S12W12 & 10:44 & M1.3 & No & 10:58 & 561 & No & 11:24 & 421 & No & -- & No \\
308 & 20250124 & 11:40 & S20E90 & 11:38 & Backside & No & No & -- & No & 12:00 & 766 & No & -- & No \\
309 & 20250202 & 05:45 & Backside & 05:33 & Backside & No & 05:40 & 736 & No & 05:48 & 1143 & No & -- & No \\
310 & 20250205 & 07:53 & N19W32 & 07:44 & M2.7 & No & 07:51 & 390 & No & 08:24 & 700 & No & -- & No \\
311 & 20250223 & 19:25 & N25W86 & 19:22 & X2.0 & No & No & -- & No & 20:06 & 1388 & No & -- & No \\
312 & 20250224 & 07:00 & N16W20 & 06:53 & M3.3 & No & 07:03 & 677 & 01 & 07:53 & 577 & No & -- & No \\
313 & 20250225 & 11:28 & S15W55 & 11:20 & M3.6 & No & 11:45 & 630 & No & 12:23 & 1346 & No & -- & No \\
314 & 20250301 & 22:53 & S20E90 & 22:44 & Backside & No & No & -- & No & 23:24 & 368 & No & -- & No \\
315 & 20250328 & 15:05 & N06E83 & 15:03 & X1.1 & No & No & -- & No & 15:24 & 1986 & No & -- & No \\

\bottomrule
\end{tabular}
\end{sidewaystable}

\begin{sidewaystable}
\centering
\scriptsize
\setlength{\tabcolsep}{5pt}
%\bfseries
\caption*{ Table \ref{table2} (continued).}
\begin{tabular}{ccccccccccccccc}
\toprule
 Event & Date & EUV & EUV & Flare & GOES & Moreton & m-type II & m-type II & SEP & CME &CME &ICME & ICME & $ |Dst|$ \\
 No. & YYYYMMDD & Onset & Source & Onset & Flare & Wave & Onset & Speed & Association/ & Onset &Speed & Disturbance & Speed & Index \\
 &      & Time (UT)  & location & Time (UT) & Class & (Yes/No) &  &   ({$\mathrm{km\,s^{-1}}$}) & Peak Flux (pfu)  & & ($\mathrm{km\,s^{-1}}$) & Date (UT) & ({$\mathrm{km\,s^{-1}}$}) & (nT) \\ 
\midrule
316 & 20250418 & 23:19 & S26E100 & 23:18 & M4.5 & No & 23:45 & 349 & No & 23:48 & 1050 & No & -- & No \\
317 & 20250420 & 11:38 & S20E85 & 11:37 & Backside & No & No & -- & No & 12:12 & 240 & No & -- & No \\
318 & 20250422 & 07:52 & S32E11 & 07:40 & C2.9 & No & No & -- & No & 08:48 & 1042 & No & -- & No \\
319 & 20250513 & 15:22 & N07W83 & 15:19 & X1.2 & No & 15:42 & 658 & 05 & 15:48 & 762 & No & -- & No \\
320 & 20250514 & 03:05 & N17E69 & 03:05 & M5.3 & No & 03:25 & 228 & No & 03:36 & 898 & No & -- & No \\
321 & 20250514 & 08:09 & N17E65 & 08:07 & X2.7 & No & No & -- & No & 08:36 & 933 & No & -- & No \\
322 & 20250514 & 18:03 & N17E60 & 17:59 & M4.7 & No & No & -- & No & 18:36 & 483 & No & -- & No \\
323 & 20250530 & 06:05 & N06E21 & 05:56 & M3.4 & No & 06:12 & 538 & No & 06:36 & 353 & 20250601 05:40 & 1100 & 119 \\
324 & 20250530 & 23:41 & N14E10 & 23:31 & M8.1 & No & 23:53 & 1938 & No & 00:12 & 1752 & No & -- & No \\
325 & 20250602 & 21:35 & N08W31 & 21:35 & M1.1 & No & No & -- & No & 22:12 & 406 & No & -- & No \\
326 & 20250615 & 07:45 & S14W67 & 07:45 & M2.2 & No & 07:56 & 488 & No & 08:12 & 510 & No & -- & No \\
327 & 20250615 & 17:55 & N17E15 & 17:45 & M8.4 & No & 18:19 & 397 & No & 18:24 & 1092 & No & -- & No \\
328 & 20250618 & 04:35 & S12W113 & 04:30 & Backside & No & 04:40 & 1471 & No & 05:00 & 574 & No & -- & No \\
329 & 20250708 & 03:47 & N20E88 & 03:42 & M2.4 & No & No & -- & No & 04:24 & 581 & No & -- & No \\
330 & 20250715 & 10:38 & S15W119 & 10:30 & Backside & No & 10:52 & 525 & No & 11:00 & 496 & No & -- & No \\
331 & 20250716 & 16:51 & S05W90 & 16:40 & Backside & No & No & -- & No & 17:24 & 343 & No & -- & No \\
332 & 20250803 & 13:42 & N03E23 & 13:40 & M2.9 & No & No & -- & No & 15:24 & 398 & No & -- & No \\
333 & 20250804 & 04:57 & N04E15 & 04:52 & M2.0 & No & No & -- & No & 05:36 & 240 & No & -- & No \\
334 & 20250805 & 15:53 & N04W07 & 15:46 & M4.4 & No & 15:51 & 734 & No & 16:36 & 398 & No & -- & No \\
335 & 20250821 & 07:58 & N12E138 & 07:55 & Backside & No & 08:22 & 1323 & No & 08:24 & 1612 & No & -- & No \\
336 & 20250830 & 19:20 & N04E07 & 19:11 & M2.7 & No & No & -- & No & 20:12 & 914 & 20250901 20:58 & 600 & 78 \\
337 & 20251002 & 23:06 & S10E90 & 23:00 & Backside & No & No & -- & No & 23:36 & 471 & No & -- & No \\
338 & 20251028 & 23:43 & N25E90 & 23:40 & Backside & No & No & -- & No & 00:12 & 1158 & No & -- & No \\
339 & 20251030 & 03:54 & N20E90 & 03:50 & Backside & No & 04:28 & 691 & No & 04:25 & 1436 & No & -- & No \\
340 & 20251031 & 19:49 & N20E90 & 19:37 & C7.1 & No & 20:13 & 542 & No & 20:24 & 785 & No & -- & No \\
341 & 20251103 & 09:00 & N22E84 & 08:46 & M5.0 & Yes & No & -- & No & 09:36 & 474 & No & -- & No \\
342 & 20251104 & 17:21 & N23E62 & 17:20 & X1.8 & No & No & -- & No & 17:36 & 1070 & No & -- & No \\
343 & 20251104 & 22:32 & N23E36 & 22:31 & M1.7 & No & No & -- & No & 23:00 & 402 & 20251107 05:16 & 710 & 58 \\
344 & 20251105 & 10:33 & N21E48 & 10:33 & M7.4 & No & No & -- & No & 11:00 & 1249 & 20251107 05:16 & 710 & 58 \\
345 & 20251105 & 21:54 & N27E40 & 21:52 & M8.6 & No & 21:59 & 1395 & No & 22:24 & 920 & 20251107 05:16 & 710 & 58 \\
346 & 20251107 & 06:47 & N23E28 & 06:33 & M1.7 & No & 07:37 & 1169 & No & 07:24 & 415 & No & -- & No \\
347 & 20251109 & 07:03 & N22E03 & 07:01 & X1.8 & No & 07:12 & 804 & No & 07:36 & 792 & 20251112 00:09 & 760 & 217 \\
348. & 20251110 & 09:07 & N22W14 & 08:55 & X1.2 & Yes & 09:11 & 1321 & 30 & 09:36 & 947 & 20251112 00:09 & 760 & 217 \\
349 & 20251111 & 09:52 & N23W23 & 09:49 & X5.1 & Yes & 10:01 & 1350 & 1460 & 10:24 & 1772 & 20251112 19:17 & 690 & 143 \\
350 & 20251114 & 07:35 & N23W62 & 07:35 & X4.0 & Yes & 08:29 & 1269 & 16 & 08:00 & 1308 & No & -- & No \\
351 & 20251201 & 02:26 & N20E81 & 02:25 & X1.9 & Yes & 02:43 & 988 & No & 02:48 & 1810 & No & -- & No \\
352 & 20251206 & 18:12 & N23E10 & 18:58 & M1.1 & No & No & -- & No & 19:48 & 228 & No & -- & No \\
353 & 20251206 & 20:29 & N20E01 & 20:29 & M8.1 & No & 20:40 & 1143 & 0.9 & 21:00 & 804 & No & -- & No \\
354 & 20251208 & 04:47 & S14W51 & 04:47 & X1.1 & No & 05:13 & 347 & No & 05:12 & 740 & 20251210 16:46 & 420 & 82 \\
355 & 20251210 & 21:39 & S19W65 & 21:38 & M4.4 & No & 22:08 & -- & No & 22:24 & 220 & No & -- & No \\
356 & 20251228 & 22:15 & N08W31 & 22:01 & M4.2 & No & No & -- & No & 23:00 & 361 & No & -- & No \\
357 & 20251231 & 13:25 & N25E20 & 13:12 & M7.1 & No & No & -- & No & 14:00 & 663 & No & -- & No \\
\bottomrule

\end{tabular}
%\end{table}
\end{sidewaystable}
%\bibliographystyle{apj}
%\bibliography{reference}
%\bibitem{latexcompanion}
%Clark D. H., Caswell J. L. 1976, MNRAS, 174, 267

%\end{theunbibliography}

\end{document}